\documentclass[twoside,english,3p]{elsarticle}

\usepackage[T1]{fontenc}
\usepackage[utf8]{inputenc}
\usepackage{babel}
\usepackage{amsmath}
\usepackage{amssymb}
\usepackage{graphicx}
\usepackage{esint}
\usepackage[unicode=true,
 bookmarks=true,bookmarksnumbered=false,bookmarksopen=false,
 breaklinks=false,pdfborder={0 0 0},backref=false,colorlinks=false]
 {hyperref}

\makeatletter

\providecommand{\tabularnewline}{\\}
\newcommand{\lyxdot}{.}

\journal{Computer Physics Communications}
\biboptions{sort&compress}

\makeatother

\begin{document}

\begin{frontmatter}{}

\title{Fourth order real space solver for the time-dependent Schrödinger
equation with singular Coulomb potential}

\author[msz]{Szilárd Majorosi\corref{cor1}}

\ead{majorosi.szilard@physx.u-szeged.hu}

\author[msz,cza]{Attila Czirják}

\cortext[cor1]{Corresponding author. }

\address[msz]{Department of Theoretical Physics, University of Szeged, Tisza L.
krt. 84-86, H-6720 Szeged, Hungary}

\address[cza]{ELI-ALPS, ELI-HU Non-Profit Ltd., Dugonics tér 13, H-6720 Szeged,
Hungary }
\begin{abstract}
We present a novel numerical method and algorithm for the solution
of the 3D axially symmetric time-dependent Schrödinger equation in
cylindrical coordinates, involving singular Coulomb potential terms
besides a smooth time-dependent potential. We use fourth order finite
difference real space discretization, with special formulae for the
arising Neumann and Robin boundary conditions along the symmetry axis.
Our propagation algorithm is based on merging the method of the split-operator
approximation of the exponential operator with the implicit equations
of second order cylindrical 2D Crank-Nicolson scheme. We call this
method hybrid splitting scheme because it inherits both the speed
of the split step finite difference schemes and the robustness of
the full Crank-Nicolson scheme. Based on a thorough error analysis,
we verified both the fourth order accuracy of the spatial discretization
in the optimal spatial step size range, and the fourth order scaling
with the time step in the case of proper high order expressions of
the split-operator. We demonstrate the performance and high accuracy
of our hybrid splitting scheme by simulating optical tunneling from
a hydrogen atom due to a few-cycle laser pulse with linear polarization.\end{abstract}

\end{frontmatter}{}

\section{Introduction\label{sec:introduction}}

Experiments with attosecond light pulses \cite{hentschel2001attosecond,Paul_Science_2001_Atto_pulsetrain,kienberger2002attosecond,drescher2002attosecond,baltuska2003attosecond,Krausz_RevModPhys_2009_Attosecond_physics},
based on high-order harmonic generation (HHG) in noble gases \cite{McPherson_JOSAB_1987_HHG,Ferray_JPhysB_1988_HHG,Farkas_PhysLettA_1992_AttoPulse,Harris_OptCom_1993_HHG_Atto},
have been revolutionizing our view of fundamental atomic, molecular
and solid state processes in this time domain \cite{Uiberacker_Nature_2007,Haessler_NatPhys_2010,pfeiffer2012ionizationmomentum,Shafir_Nature_2012,Schultze_Science_2010,Cavalieri_Nature_2007,BOOK_QUANTUM_IMAGING_BANDRAUK_2011}.
A key step in gas-HHG is the tunnel ionization of a single atom and
the return of the just liberated electron to its parent ion \cite{Keldysh_JETP_1965,Corkum_PRL_1993},
due to the strong, linearly polarized femtosecond laser pulse driving
this process. Recent developments in attosecond physics revealed that
the accurate description of this single-atom emission is more important
than ever, e.g. for the correct interpretation of the data measured
in attosecond metrology experiments, especially regarding the problem
of the zero of time \cite{eckle2008attosecond,Schultze_Science_2010},
and the problem of the exit momentum \cite{pfeiffer2012ionizationmomentum}.
Although an intuitive and very successful approximate analytical solution
\cite{Lewenstein_PRA_1994} and many of its refinements exist \cite{ivanov2005strongfield},
the most accurate description of the single-atom response is given
by the numerical solution of the time-dependent Schrödinger equation
(TDSE).

The common model of the single-atom response employs the single active
electron approximation and the dipole approximation. These reduce
the problem to the motion of an electron in the time-dependent potential
formed by adding the effect of the time-dependent electric field to
the atomic binding potential. The peculiarity of this problem is due
to the strength of the electric field, which has its maximum typically
in the range of 0.05-0.1 atomic units. Thus, this electric field enables
the tunneling of the electron through the (time-dependent) potential
barrier formed by strongly distorting the atomic potential, but this
effect is weak during the whole process. On the other hand, this small
part of the wave function outside the barrier extends to large distances
and in fact this is the main contribution to the time-dependent dipole
moment, which is usually considered as the source of the emitted radiation.
Thus, a very weak effect needs to be computed very accurately, and
these requirements get even more severe, if the model goes beyond
the single active electron or the dipole approximation.

The fundamental importance of the singularity of the Coulomb potential
regarding HHG has already been analysed and emphasized e.g. in \cite{gordon2005tdsecoulomb}.
As other kind of numerical errors decrease, the artifacts caused by
the smoothing of the singularity become more disturbing. Therefore,
a high precision numerical method has to handle the Coulomb singularity
as accurately and effectively as possible.

The most widely employed approach to the numerical solution of the
TDSE for an atomic electron driven by a strong laser pulse is to use
spherical polar coordinates: then the hydrogen eigenfunctions are
analytic and one expands the wavefunction in $Y_{l,m}(\theta,\varphi)$
spherical harmonics with expansion coefficients $\phi_{l,m}(r,t)$
acting as radial or reduced radial functions, for which the problem
is solved directly. For example, these radial functions can be further
expanded using a suitable basis, like Gaussian-Legendre or B-splines
\cite{tong1997tdsehhgspectral,bachau2001tdsebspline,cormier1997tdsebspline},
in which the Hamiltonian matrix does not have any singularity. Also,
refined solutions based on highly accurate real space discretization
of the radial coordinate $r$ exist using this harmonic expansion
\cite{muller1999tdseefficient,bauer2006tdseqprop,smyth1998tdsehelium,argenti2013photoionization,schneider2006tdsesolver},
also relying on the fact that a boundary condition for $\phi_{l,m}(0,t)$
can be derived from the analytic properties of hydrogenic eigenfunctions
and can be built into the implicit equations. The usual drawback is
that in order to calculate physical quantities either the real space
reconstruction of the wavefunction is necessary or every observable
must be given in this harmonic expansion. Also, the spherical grid
of these methods is not well adapted to the electron's motion which
is mainly along the polarization direction of the laser field.

One can easily stumble upon applications other than HHG where neither
spherical coordinates, nor special basis functions are optimal, and
one has to use Cartesian or cylindrical coordinate systems in numerical
simulations, and of course, discretization of real space coordinates.
The most basic method is just to use a staggered grid which avoids
the singularity of the Coulomb potential \cite{kolakowska1998excitation}:
then the scheme will be easily solvable, although with low accuracy.
A somewhat refined version of this is using a smoothed version of
the potential, for example the so called soft Coulomb-form of $1/\sqrt{\alpha^{2}+|{\bf r}|^{2}}$,
where the value of $\alpha$ can be fitted by matching the ground
state energy with that of the true hydrogen atom \cite{kolakowska1998excitation,christov1997tdsehhg,hu2004tdsehhg,wells1996tdsenumerical}.
These approximations are rather crude, but can reproduce important
features of the quantum system \cite{gordon2005tdsecoulomb}. Recently,
Gordon et. al. \cite{gordon2006tdsecoulomb} provided an improved
way to include the Coulomb singularity, by using the general asymptotic
behavior of the hydrogen eigenstates around $|{\bf r}|=0$. They built
an improved discrete Hamiltonian matrix through a new discrete potential,
using a three point finite difference scheme. This way they increased
the accuracy of the numerical solution by an order of magnitude.

In the present work, we propose a novel method to accurately incorporate
the effect of the Coulomb singularity in the solution of a spinless,
axially symmetric TDSE, in cylindrical coordinates. We account for
the singular Coulomb potential through analytic boundary condition
which resolves the problem of nondifferentiability of the hydrogen
eigenfunctions in the cylindrical coordinate system and yields high
order spatial accuracy.

Incorporating this kind of analytic boundary condition into the numerical
solution of TDSE will constrain the applicable propagation methods
because the singularity of the Coulomb potential at $r=0$ and the
singularity of the cylindrical radial coordinate at $\rho=0$ will
introduce Robin and Neumann boundary conditions, overriding the Hamilton
operator. As a consequence, typical explicit methods, for example
staggered leapfrog \cite{gordon2006tdsecoulomb,gainullin2015tdsesolver},
second order symmetric difference method \cite{leforestier1991tdsecomparison,iitaka1994tdsemsd4},
any polynomial expansion method \cite{leforestier1991tdsecomparison,castro2004kohnshampropagators},
and split-step Fourier transformation methods \cite{leforestier1991tdsecomparison,blanes2000tdsesplitting,chin2001splittinggrad}
have been ruled out, leaving only implicit methods.

In Section \ref{sec:introduction_main}, we introduce the cylindrical
TDSE to be solved, along with its boundary conditions, and we describe
the way how we incorporate the singular Coulomb potential to the problem.
In Section \ref{sec:crank_nicolson_main}, we derive the suitable
finite-difference scheme with high order spatial accuracy, using second
order Crank-Nicolson method \cite{leforestier1991tdsecomparison,BOOK_NUMERICAL_RECIPIES_2007,puzynin1999tdsepade}.
We briefly go through the standard approximations of the evolution
operator, then we write down the detailed form of the resulting implicit
linear equations. In Section \ref{sec:operator_splitting_main}, we
overview the standard split-operator methods for numerically solving
the TDSE, and combine them with the finite-difference description,
which is known as split-step finite difference method \cite{wang2005nltdsedifferences,shen2013tdsesolving}.
We show that similarly to the typical explicit methods, real space
split-operator methods do not work well for singular Hamiltonians.
Based in these conclusions, we introduce our hybrid splitting method,
which retains the robustness of the full Crank-Nicolson method and
the speed of the split-operator method. In Section \ref{sec:hybrid_splitting_algorithm},
we analyse the newly arisen coefficient matrix, we describe the way
how to take advantage of its structure, then we derive a special solver
algorithm which is necessary for fast solution of the TDSE. In Section
\ref{sec:numerical_results_main}, we discuss the results of numerical
experiments we carried out with our hybrid splitting method: (i) for
the accuracy of the spatial discretization of the finite difference
scheme, we calculate the eigenenergies of selected eigenstates, (ii)
for the temporal accuracy of our split-step finite difference scheme
combined with high order evolution operator factorizations, we compare
the numerical wave function with the analytic solution of the quantum
forced harmonic oscillator, (iii) for final verification, we simulate
the hydrogen atom in a time dependent external electric field.

\section{Schrödinger equation and boundary conditions\label{sec:introduction_main}}

\subsection{The time-dependent Schrödinger equation\label{sub:tdse_3dc}}

Let us consider the axially symmetric three-dimensional time-dependent
Schrödinger equation in cylindrical coordinates $\rho=\sqrt{x^{2}+y^{2}}$
and $z$: 
\begin{equation}
i\hbar\frac{\partial}{\partial t}\Psi\left(z,\rho,t\right)=-\frac{\hbar^{2}}{2\mu}\left[\frac{\partial^{2}}{\partial z^{2}}+\frac{\partial^{2}}{\partial\rho^{2}}+\frac{1}{\rho}\frac{\partial}{\partial\rho}\right]\Psi\left(z,\rho,t\right)+V\left(z,\rho,t\right)\Psi\left(z,\rho,t\right)\label{eq:tdse_3dc}
\end{equation}
where $\mu$ is the (reduced) electron mass. For the sake of simplicity,
we use atomic units throughout this article and we use the notation
$\beta=-\hbar^{2}/2\mu$.

As discussed in the Introduction, such a Hamiltonian plays a fundamental
role e.g. in the interaction of atomic systems with strong laser pulses.
Then the potential is usually the sum of an atomic binding potential
(centered in the origin) and an interaction energy term. For a single
active electron and a linearly polarized laser pulse, this interaction
term in dipole approximation and in length gauge \cite{BOOK_ATOMS_MOLECULES_BRANSDEN_2003}
reads as:

\begin{equation}
V\left(z,\rho,t\right)=V_{\text{atom}}\left(z,\rho\right)-q\cdot F(t)\cdot z,\label{eq:tdse_pot_dipole}
\end{equation}
where $q=-1$ denotes the electron’s electric charge and $F(t)$ is
the laser's electric field, pointing in the $z$ direction.

The wave function is defined within the interval $z\in\left[z_{\min},z_{\max}\right]$,
$\rho\in\left[0,\rho_{\max}\right]$ and the problem is treated as
an initial value problem as $\Psi\left(z,\rho,0\right)=\psi\left(z,\rho\right)$
is known. Box boundary conditions are posed at three edges of the
interval: 
\begin{equation}
\Psi\left(z>z_{\max},\rho,t\right)=\Psi\left(z<z_{\min},\rho,t\right)=\Psi\left(z,\left|\rho\right|>\rho_{\text{max}},t\right)=0\label{eq:tdse_bc_box}
\end{equation}
and because of the singularity in the curvilinear coordinate $\rho$,
Neumann boundary condition should be posed along the line $\rho=0$.
In order to find this boundary condition, we multiply (\ref{eq:tdse_3dc})
by $\rho$ and take the limit $\rho\rightarrow0$: 
\begin{equation}
\left.\frac{\partial\Psi}{\partial\rho}\right|_{\rho=0}=-\frac{1}{\beta}\lim_{\rho\rightarrow0}\left(\rho V\left(z,\rho,t\right)\Psi\left(z,\rho,t\right)\right)=0\label{eq:tdse_bc_nm}
\end{equation}
where we have assumed the potential is smooth at $\rho=0$ line. Then
$\Psi(z,\rho,t)$ is continuous and continuously differentiable everywhere
in the $z-\rho$ plane, and it has extremal value in $\rho=0$. Symmetry
condition $\Psi(z,\rho,t)=\Psi(z,-\rho,t)$ also holds in this case.

However, this Neumann boundary condition has to be altered when we
introduce a singular potential in any form, e.g. the 3D Coulomb potential
$V_{{\rm atom}}\left(z,\rho\right)=\gamma/r$ with $r=\sqrt{z^{2}+\rho^{2}}$
and $\gamma=-q^{2}Z/4\pi\epsilon_{0}$.

\subsection{The boundary condition for the 3D Coulomb potential\label{sec:tdse_3d_coulomb}}

Our aim is to represent the singularity of the 3D Coulomb potential
by implementing the correct boundary condition in the origin. In order
to find this condition, it seems necessary to use spherical polar
coordinates, since the Coulomb problem is not separable in cylindrical
coordinates. Let us first discuss the properties of the radial solutions
$\phi(r)$ of the well known radial equation \cite{BOOK_QUANTUM_COHEN_1997,BOOK_QUANTUM_GRIFFITS_2005}:
\begin{equation}
\left[\beta\frac{\partial^{2}}{\partial r^{2}}+\frac{2\beta}{r}\frac{\partial}{\partial r}+\beta\frac{l(l+1)}{r^{2}}+\frac{\gamma}{r}\right]\phi(r)=E\phi(r),\label{eq:tdse_3d_rad}
\end{equation}
where $l$ is the angular momentum quantum number. If we multiply
both sides with $r$ then take the $r\rightarrow0$ limit, we can
acquire Robin boundary condition for all of the $l=0$ states as 
\begin{equation}
\left.\frac{\partial\phi}{\partial r}\right|_{r=0}=-\frac{\gamma}{2\beta}\phi\left(0\right),\label{eq:tdse_3d_rad_bc_l}
\end{equation}
but the bound states with higher angular momenta will have $\phi\left(0\right)=0$
boundary condition at the origin, in accordance with the fact that
for $l>0$ the particle physically cannot penetrate to $r=0$, because
of the singularity of $l(l+1)/r^{2}$.

Since we need the boundary conditions for the cylindrical equation
independent of the $l$ values, we turn to the nonseparated symmetric
spherical Schrödinger equation for answers:

\begin{equation}
\left[\beta\frac{\partial^{2}}{\partial r^{2}}+\frac{2\beta}{r}\frac{\partial}{\partial r}+\beta\frac{1}{r^{2}\sin\theta}\frac{\partial}{\partial\theta}\left(\sin\theta\frac{\partial}{\partial\theta}\right)+\frac{\gamma}{r}\right]\Psi(r,\theta)=E\Psi(r,\theta)\label{eq:tdse_3d_rad_theta}
\end{equation}
which has also the term $1/r^{2}$ and has singularity at $r=0$ and
at $\theta=k\pi$, $k=0,1,2\dots$. To acquire the boundary conditions
for $r=0$, we again multiply both sides with $r$ and take the $r\rightarrow0$
limit, yielding 
\begin{equation}
\lim_{r\rightarrow0}\left[2\beta\frac{\partial}{\partial r}+\beta\frac{\cos\theta}{r\sin\theta}\frac{\partial}{\partial\theta}+\frac{\beta}{r}\frac{\partial^{2}}{\partial^{2}\theta}+\gamma\right]\Psi(r,\theta)=0.\label{eq:tdse_coulomb_spherical}
\end{equation}
Now we transform equation (\ref{eq:tdse_coulomb_spherical}) into
cylindrical coordinates $z=r\cos\theta$ and $\rho=r\sin\theta$,
using following expressions of the partial derivatives:

\begin{equation}
\frac{\partial}{\partial r}=\frac{\partial z}{\partial r}\frac{\partial}{\partial z}+\frac{\partial\rho}{\partial r}\frac{\partial}{\partial\rho}=\sin\theta\frac{\partial}{\partial\rho}+\cos\theta\frac{\partial}{\partial z},
\end{equation}

\begin{equation}
\frac{\partial}{\partial\theta}=\frac{\partial z}{\partial\theta}\frac{\partial}{\partial z}+\frac{\partial\rho}{\partial\theta}\frac{\partial}{\partial\rho}=r\cos\theta\frac{\partial}{\partial\rho}-r\sin\theta\frac{\partial}{\partial z}
\end{equation}
with $\cos\theta=z/\sqrt{z^{2}+\rho^{2}}$ and $\sin\theta=\rho/\sqrt{z^{2}+\rho^{2}}$.
After writing back and taking the limit, the formula (\ref{eq:tdse_coulomb_spherical})
becomes

\begin{equation}
\left.\left[2\beta\frac{\rho}{\sqrt{z^{2}+\rho^{2}}}\frac{\partial}{\partial\rho}+\beta\frac{z}{\sqrt{z^{2}+\rho^{2}}}\frac{\partial}{\partial z}+\beta\frac{z^{2}}{\rho\sqrt{z^{2}+\rho^{2}}}\frac{\partial}{\partial\rho}+\gamma\right]\Psi(r,\theta)\right|_{r=0}=0.\label{eq:tdse_3d_rad_bc_nm}
\end{equation}
Then, by substituting $z=0$ into (\ref{eq:tdse_3d_rad_bc_nm}), we
obtain the Robin boundary condition for the 3D Coulomb singularity:
\begin{equation}
\left.\frac{\partial\Psi}{\partial\rho}\right|_{\rho,z=0}=-\frac{\gamma}{2\beta}\Psi\left(0,0,t\right).\label{eq:tdse_3d_bc_coulomb}
\end{equation}
This can be generalized to include multiple Coulomb-cores rather easily,
we just need to impose (\ref{eq:tdse_3d_bc_coulomb}) at multiple
$z,\rho$ points along the $\rho=0$ axis. Additionally, any continuously
differentiable potential added to this configuration does not change
this boundary condition. It is interesting to note also that the boundary
condition imposed by a 1D Dirac-delta potential $\gamma\delta(\rho)$
also has the form of (\ref{eq:tdse_3d_bc_coulomb}) for a symmetric
wave function \cite{BOOK_QUANTUM_GRIFFITS_2005}.

\subsection{The states with nonzero magnetic quantum numbers\label{sec:tdse_3d_state_m}}

Although we are mainly interested in the solution of the TDSE with
an axially symmetric initial state, i.e. an initial state of zero
magnetic quantum number $m$, we briefly show in the following that
initial states with nonzero $m$ also lead to a TDSE of the form (\ref{eq:tdse_3dc_full_zr}),
however with different boundary conditions at $\rho=0$.

Let us write out the full 3D cylindrical time-dependent Schrödinger
equation, again with the same axially symmetric potential that we
used previously: 
\begin{equation}
i\frac{\partial}{\partial t}\Xi\left(z,\rho,\phi,t\right)=\left[\beta\frac{\partial^{2}}{\partial z^{2}}+\beta\frac{\partial^{2}}{\partial\rho^{2}}+\frac{\beta}{\rho}\frac{\partial}{\partial\rho}+\frac{\beta}{\rho^{2}}\frac{\partial^{2}}{\partial\phi^{2}}+V\left(z,\rho,t\right)\right]\Xi\left(z,\rho,\phi,t\right).\label{eq:tdse_3dc_full_zr}
\end{equation}
We can write the quantum state with a given magnetic quantum number
$m$ in the following form for any $t$

\begin{equation}
\Xi\left(z,\rho,\phi,t\right)=\Psi_{m}(z,\rho,t)e^{im\phi},\label{eq:tdse_3dc_state_m}
\end{equation}
due to the fact that the Hamiltonian of (\ref{eq:tdse_3dc_full_zr})
conserves the $m$ quantum number. (We can also use real angular basis
functions like $\sin(m\phi)$ or $\cos(m\phi)$ instead of $e^{im\phi}$,
as long as that they are the eigenfunctions of $\partial_{\phi}^{2}$.)

After substituting (\ref{eq:tdse_3dc_state_m}) to (\ref{eq:tdse_3dc_full_zr})
we perform projection in the form of $\intop_{0}^{2\pi}e^{-im\phi}(\cdot){\rm d}\phi$
. Then we get the TDSE for our wave function $\Psi_{m}(z,\rho,t)$
with the specified magnetic quantum number $m$:

\begin{equation}
i\frac{\partial}{\partial t}\Psi_{m}(z,\rho,t)=\left[\beta\frac{\partial^{2}}{\partial z^{2}}+\beta\frac{\partial^{2}}{\partial\rho^{2}}+\frac{\beta}{\rho}\frac{\partial}{\partial\rho}-\beta\frac{m^{2}}{\rho^{2}}+V\left(z,\rho,t\right)\right]\Psi_{m}(z,\rho,t),\label{eq:tdse_3dc_full_m}
\end{equation}
Eq. (\ref{eq:tdse_3dc_full_m}) has the same form as the axially symmetric
equation (\ref{eq:tdse_3dc}) but with a new $m$ dependent potential:
\begin{equation}
V_{m}(z,\rho,t)=V\left(z,\rho,t\right)-\beta m^{2}/\rho^{2}.\label{eq:tdse_3dc_pot_m}
\end{equation}
However, if $m\neq0$ then the $1/\rho^{2}$ singularity of (\ref{eq:tdse_3dc_pot_m})
alters the boundary condition (\ref{eq:tdse_bc_nm}) at the $\rho=0$
axis to the following: 
\begin{equation}
\Psi_{m}(z,0,t)=0,\label{eq:tdse_3dc_bc_m}
\end{equation}
which can by derived by multiplying (\ref{eq:tdse_3dc_full_m}) with
$\rho^{2}$ and taking the limit of $\rho\rightarrow0$. The boundary
condition (\ref{eq:tdse_3dc_bc_m}) also works with Coulomb potential,
and in every case where $\rho^{2}V\left(z,\rho,t\right)=0$ in the
limit of $\rho\rightarrow0$. It is also consistent with the boundary
conditions for Coulomb states with nonzero angular momenta ($l\neq0$)
mentioned in the previous section.

\section{The Crank-Nicolson method\label{sec:crank_nicolson_main}}

\subsection{Approximation of the time evolution operator\label{sec:time_evolution}}

The formal solution of (\ref{eq:tdse_3dc}) in terms of the time evolution
operator $U(t,t')=\mathcal{T}\exp\left[-\frac{i}{\hbar}\int H(t'')dt''\right]$
is the following

\begin{equation}
|\Psi(t)\rangle=U(t,t')|\Psi(t')\rangle.
\end{equation}
Here, the exponential operator is to be understood as a time-ordered
quantity, which is a difficult procedure if the Hamilton operators
at different time instants do not commute. However, this is the case
for the potential we are considering.

To acquire suitable discretization in the time domain of problem (\ref{eq:tdse_3dc})
we approximate the $U(t,0)$ time evolution operator directly. First,
let us divide the time domain $[0,t]$, into $N_{t}$ equal subintervals,
then by the group property of $U(t,0)$ we get

\begin{equation}
U\left(t,0\right)=\prod_{k=0}^{N_{t}-1}U\left(t_{k+1},t_{k}\right)\label{eq:time_u}
\end{equation}
where $t_{k}=k\Delta t$ and $\Delta t=t/N_{t}$. In one interval,
we can write the evolution operator with its short time form of 
\begin{equation}
U\left(t_{k+1},t_{k}\right)=e^{-i\Delta t\, H_{k}}\label{eq:time_u_short}
\end{equation}
where $H_{k}$ is the effective time-independent Hamiltonian related
to the original one $H(t)$ by the Magnus expansion \cite{puzynin2000tdsemagnus,blanes2009magnus}:
\begin{equation}
H_{k}=\frac{1}{\Delta t}\intop_{t_{k}}^{t_{k+1}}H(t')\text{d}t'+\frac{i}{2\Delta t}\intop_{t_{k}}^{t_{k+1}}\intop_{t_{k}}^{t^{'}}\left[H(t''),H(t')\right]\text{d}t''\text{d}t'+\dots.\label{eq:time_h_magnus}
\end{equation}
This includes infinitely many commutators of the Hamiltonians evaluated
at different time points, to be integrated with respect to more and
more variables.

Using only the first term of this series, one can directly acquire
the well known second order approximation of the Hamiltonian as 
\begin{equation}
H_{k}^{(2)}=H(t_{k+\frac{1}{2}}).\label{eq:time_h_eff_2}
\end{equation}
In order to have information about the next error term of the time
evolution, we need to evaluate the Magnus commutator series to fourth
order. According to Puzynin et. al. \cite{puzynin2000tdsemagnus},
this improved approximation for a TDSE of the form (\ref{eq:tdse_3dc})
is 
\begin{equation}
H_{k}^{(4)}=\frac{1}{2\mu}\left(-i\nabla+\frac{\Delta t^{2}}{12}\nabla\dot{V}(t_{k+1/2})\right)^{2}+V(t_{k+1/2})+\frac{\Delta t^{2}}{24}\ddot{V}(t_{k+1/2}).\label{eq:time_h_eff_4}
\end{equation}
where the top dots are the short hand notation for $\partial_{t}$
time derivatives. This formula is important even if we use only the
second order approximation, because the leading order of error depends
on characteristics of the first and second time derivatives of $V(z,\rho,t)$.

For the exponential operators (\ref{eq:time_u_short}), first we consider
the diagonal Padé-approximation of an exponential function 
\begin{equation}
e^{\lambda\cdot x}=\frac{_{1}F_{1}(-M,-2M,\lambda\cdot x)}{_{1}F_{1}(-M,-2M,-\lambda\cdot x)}+O(\lambda^{2M+1})\label{eq:time_pade_diagonal}
\end{equation}
where $_{1}F_{1}$ is the confluent hypergeometric function, which
in this case reduces to a polynomial of degree $M$ with real coefficients.
This expression can be used for the exponential operators (\ref{eq:time_u_short})
with $\lambda=-i\Delta t$, and it can be shown that for self-adjoint
operators the approximation is unitary \cite{gordon2005tdsecoulomb}.

From this, a generalized operator approximation scheme that is $\Delta t^{2M+1}$
accurate can be obtained \cite{puzynin1999tdsepade,dijk2007tdseaccurate}
in a general form of 
\begin{equation}
e^{-i\Delta t\, H_{k}}=\prod_{s=1}^{M}\left[\left(1+i\frac{\Delta t}{x_{s}^{*}}H_{k}\right)^{-1}\left(1-i\frac{\Delta t}{x_{s}}H_{k}\right)\right]+O(\Delta t^{2M+1})\label{eq:time_pade_cn_gen}
\end{equation}
where $x_{s}$ for $s=1,\dots,M$ are the roots of the polynomial
equation 
\begin{equation}
_{1}F_{1}\left(-M,-2M,-x\right)=0.\label{eq:time_pade_cn_gen_f}
\end{equation}
If we truncate both the Magnus-series at the first term in (\ref{eq:time_h_magnus})
and take a single coefficient of the Padé-approximation ($M=1$),
then we arrive at the second order accurate implicit Crank-Nicolson
scheme: 
\begin{equation}
\Psi(t_{k+1})=\left(1+iH_{k}\frac{\Delta t}{2}\right)^{-1}\left(1-iH_{k}\frac{\Delta t}{2}\right)\Psi(t_{k}).\label{eq:time_pade_cn_2}
\end{equation}
One can straightforwardly construct higher order Crank-Nicolson schemes
using (\ref{eq:time_pade_cn_gen}), yielding multiple implicit Crank-Nicolson
substeps \cite{puzynin1999tdsepade}. For time dependent Hamiltonians
though, a high order accurate scheme should use the corresponding
high order effective Hamiltonian (\ref{eq:time_h_magnus}) for consistency.

\subsection{The finite difference scheme\label{sec:crank_nicolson_3dc}}

Our next step is to discretize the effective Hamiltonian $H_{k}$,
and to construct the Hamiltonian matrix. We propose to use the method
of fourth order finite differences with $z,\rho$ cylindrical coordinates
and the following equidistant 2D spatial grid: 
\begin{equation}
z_{i}=z_{\min}+i\cdot\Delta z,\,\Delta z=(z_{\max}-z_{\min})/N_{z},\, i\in\left[0,N_{z}\right],\label{eq:cn_spatial_z}
\end{equation}
\begin{equation}
\rho_{j}=j\cdot\Delta\rho,\,\Delta\rho=\rho_{\max}/N_{\rho},\, j\in\left[0,N_{\rho}\right].\label{eq:cn_spatial_r}
\end{equation}

The first term of the Hamiltonian $H_{k}$ is the Laplacian $\nabla^{2}$.
We denote the finite difference approximation of its $z$- and $\rho$-dependent
terms by $L_{z}$ and $L_{\rho}$, respectively, and we use the following
fourth order accurate finite difference forms \cite{dijk2007tdseaccurate}
for them:

\begin{equation}
L_{z}\Psi_{i,j}(t)=\frac{-\Psi_{i-2,j}+16\Psi_{i-1,j}-30\Psi_{i,j}+16\Psi_{i+1,j}-\Psi_{i+2,j}}{12\Delta z^{2}},\label{eq:cn_laplacian5_z}
\end{equation}

\begin{equation}
L_{\rho}\Psi_{i,j}(t)=\frac{(-1+1/j)\Psi_{i,j-2}+(16-8/j)\Psi_{i,j-1}-30\Psi_{i,j}+(16+8/j)\Psi_{i,j+1}+(-1-1/j)\Psi_{i,j+2}}{12\Delta\rho^{2}}.\label{eq:cn_laplacian5_r}
\end{equation}
These fourth order formulae are optimal in the sense that symmetric
finite differences of more than five points are very complicated to
be applied for the Coulomb-problem, because the higher order the finite
difference formula is, the higher derivatives of $\Psi(z,\rho,t)$
should be continuous.

To discretize the Neumann and Robin boundary conditions, we need a
one-sided finite difference formula for the first derivative. Based
on the method of Ref. \cite{dijk2007tdseaccurate}, we derived the
following fourth order accurate forward difference operator:

\begin{equation}
D_{\rho}\Psi_{i,j}(t)=\frac{-25\Psi_{i,j}+48\Psi_{i,j+1}-36\Psi_{i,j+2}+16\Psi_{i,j+3}-3\Psi_{i,j+4}}{12\Delta\rho}.\label{eq:cn_diff5_r}
\end{equation}

Using standard second order accurate form (\ref{eq:time_pade_cn_2})
of the exponential operator with the discretized Laplacian in the
Hamilton matrix, we get the following implicit scheme for all $i\in\left[0,N_{z}\right]$
and $j\in\left[1,N_{\rho}\right]$:

\begin{equation}
\left(1+\alpha\beta L_{z}+\alpha\beta L_{\rho}+\alpha V_{i,j}\right)\Psi_{i,j}(t_{k+1})=\left(1-\alpha\beta L_{z}-\alpha\beta L_{\rho}-\alpha V_{i,j}\right)\Psi_{i,j}(t_{k}),\label{eq:cn_3dc_form_gen}
\end{equation}
where $\alpha=i\Delta t/2$ and the potential is evaluated at the
temporal midpoint $V_{i,j}=V(z_{i},\rho_{j},t_{k+1/2})$. The box
boundary conditions $\Psi_{i,j}=0$ is applied if $j>N_{\rho}$ or
$i>N_{z}$ or $i<0$.

Assuming that the potential is smooth, the Neumann boundary condition
(\ref{eq:tdse_bc_nm}) prescribes the following implicit equations
at $j=0$ for all $i\in[0,N_{z}]$: 
\begin{equation}
D_{\rho}\Psi_{i,0}(t_{k+1})=0.\label{eq:cn_3dc_bc_nm}
\end{equation}
On the other hand, if a Coulomb-core of strength $\gamma$ is present
at the grid point $z_{R}=0$ and $m=0$ then, according to (\ref{eq:tdse_3d_bc_coulomb}),
equation (\ref{eq:cn_3dc_bc_nm}) will be overridden at $i=R$ by

\begin{equation}
\left(D_{\rho}+\frac{\gamma}{2\beta}\right)\Psi_{R,0}(t_{k+1})=0.\label{eq:cn_3dc_bc_nm_cb}
\end{equation}
Here we have assumed that the origin is included in (\ref{eq:cn_spatial_z})
with $z_{R}=0$, where $R$ can be anywhere in the interval $[0,N_{z}]$
if it is reasonably far from the box boundaries.

For $m\neq0$ states, the equations of the boundary conditions at
$\rho=0$ are simply $\Psi_{i,j=0}(t_{k+1})=0$ with or without Coulomb
potential.

For simplicity, let us introduce the short hand notations $X_{i,j}=\Psi_{i,j}(t_{k+1})$,
$\Psi_{i,j}=\Psi_{i,j}(t_{k}),$ then by substituting the finite difference
Laplacians (\ref{eq:cn_laplacian5_z}), (\ref{eq:cn_laplacian5_r})
into (\ref{eq:cn_3dc_form_gen}), we arrive at the following final
form of linear equations for all $i\in\left[0,N_{z}\right]$, $j\in\left[1,N_{\rho}\right]$:

\[
(-1+1/j)\beta_{\rho}X_{i,j-2}+(16-8/j)\beta_{\rho}X_{i,j-1}+(16+8/j)\beta_{\rho}X_{i,j+1}+(-1-1/j)\beta_{\rho}X_{i,j+2}
\]
\[
-\beta_{z}X_{i-2,j}+16\beta_{z}X_{i-1,j}+(1-30\beta_{\rho}-30\beta_{z}+\alpha V_{i,j})X_{i,j}+16\beta_{z}X_{i+1,j}-\beta_{z}X_{i+2,j}
\]
\[
=(1-1/j)\beta_{\rho}\Psi_{i,j-2}+(-16+8/j)\beta_{\rho}\Psi_{i,j-1}+(-16-8/j)\beta_{\rho}\Psi_{i,j+1}+(1+1/j)\beta_{\rho}\Psi_{i,j+2}
\]
\begin{equation}
+\beta_{z}\Psi_{i-2,j}-16\beta_{z}\Psi_{i-1,j}+(1+30\beta_{\rho}+30\beta_{z}-\alpha V_{i,j})\Psi_{i,j}-16\beta_{z}\Psi_{i+1,j}+\beta_{z}\Psi_{i+2,j},\label{eq:cn_3dc_full}
\end{equation}
where 
\begin{equation}
\alpha=i\Delta t/2,\,\,\,\beta_{\rho}=\alpha\beta/(12\Delta\rho^{2}),\,\,\,\beta_{z}=\alpha\beta/(12\Delta z^{2}).\label{eq:cn_3dc_full_ab}
\end{equation}
For the $m=0$ configuration, the equations forced by the Neumann
and Robin boundary conditions for all $i\in\left[0,N_{z}\right]$
from (\ref{eq:cn_diff5_r}), (\ref{eq:cn_3dc_bc_nm}), (\ref{eq:cn_3dc_bc_nm_cb})
are 
\begin{equation}
(-25+(6(\gamma/\beta)\Delta\rho\cdot\delta_{R,i})X_{i,0}+48X_{i,1}-36X_{i,2}+16X_{i,3}-3X_{i,4}=0.\label{eq:cn_3dc_full_nm_cb}
\end{equation}
For the $m\neq0$ states we have simply

\begin{equation}
X_{i,0}=0.\label{eq:cn_3dc_full_bc_m}
\end{equation}

The spatial discretization in the cylindrical coordinate system and
the Neumann or Robin boundary conditions for the $m=0$ states make
the unitarity of the algorithm, and in a broader sense, the accuracy
of spatial integrations a more subtle issue than usual. One has to
find an appropriate discrete inner product formula that is conserved
at least with the accuracy of the finite differences, and which can
be evaluated with sufficient accuracy using the cylindrical grid.
We give a solution to this auxiliary problem in \ref{sub:dot_product_3dc}.

Although it corresponds to 3D propagation, we call this scheme 2D
Crank-Nicolson method because it involves only two spatial coordinates.
This is already a complete propagation algorithm by itself, however,
it suffers from the numerically inefficient solution of the resulting
linear equations: if we combine the $i,j$ indices into a single one
(by flattening the two dimensional array) as $l=i\cdot(N_{\rho}+1)+j$,
we obtain a a block pentadiagonal matrix of size $(N_{z}+1)^{2}(N_{\rho}+1)^{2}$,
with block size $(N_{\rho}+1)^{2}$. Inverting this type of matrix
is computationally intense \cite{BOOK_NUMERICAL_RECIPIES_2007,varah1972blocktridiagonal}
because the width of the diagonal is $4N_{\rho}+1$: despite its apparent
simplicity, the numerical cost of this task is $\sim N_{z}N_{\rho}^{3}$
which is extremely high compared to the cost $\sim N_{z}N_{\rho}$
of a pentadiagonal scheme of the same size. These facts inspired us
to develop an improved algorithm which has almost all advantages of
this 2D Crank-Nicolson method but needs much less numerical effort.

\section{Operator splitting schemes\label{sec:operator_splitting_main}}

As we have seen in the previous section, the 2D Crank-Nicolson scheme
with the boundary condition (\ref{eq:tdse_bc_nm}) is a possible but
ineffective way of solving the TDSE numerically. Now we are going
to discuss how to apply the well-known method of operator splitting
to the solution of the problem described in the introduction.

\subsection{Operator splitting formulae\label{sub:splitting_formulas}}

The approach of the operator splitting method is to factorize the
exponential operator $e^{\lambda H}$ into multiple easy-to-solve
parts. From the Taylor-expansion of the exponential operator 
\begin{equation}
e^{\lambda(A+B)}=\sum_{n=0}^{\infty}\frac{\left(A+B\right)^{n}}{n!}\lambda^{n},\label{eq:splitting_basic_exp}
\end{equation}
follow the two main formulae \cite{wilcox1967exponential} which form
the basis of the operator splitting schemes, namely the Baker-Campbell-Hausdorff
formula :

\begin{equation}
e^{\lambda A}e^{\lambda B}=e^{\lambda(A+B)+\lambda^{2}\frac{1}{2}[B,A]+\lambda^{3}\frac{1}{12}[A-B,[A,B]]}+O(\lambda^{4}),\label{eq:splitting_basic_baker}
\end{equation}
and the Zassenhaus formula

\begin{equation}
e^{\lambda(A+B)}=e^{\lambda A}e^{\lambda B}e^{\lambda^{2}\frac{1}{2}[A,B]}e^{\lambda^{3}\frac{1}{6}[A+2B,[A,B]]}+O(\lambda^{4}).\label{eq:splitting_basic_zassenhaus}
\end{equation}
Both of these contain infinitely many commutators of $A$ and $B$.
Of course, if $[A,B]=0$ then $e^{\lambda(A+B)}=e^{\lambda A}e^{\lambda B}$
exactly. As the above formulae suggest, the $O(\lambda^{4})$ terms
can be further factorized into the exponents. An extended analysis
is available in Refs. \cite{mclachlan2002splitting,yazici2010operator}.

If one uses (\ref{eq:splitting_basic_baker}) and (\ref{eq:splitting_basic_zassenhaus})
to acquire a symmetric decomposition, only odd leading order of $\lambda$
will appear in the formula. This is a requirement for quantum propagation
though, because the presence of even order terms would destroy the
unitary evolution of the wave function. A well-known example of this
is the widely used standard symmetric second order accurate formula
(or Strang splitting, after \cite{strang1968splitting}): 
\begin{equation}
e^{\lambda(A+B)}=e^{\lambda A/2}e^{\lambda B}e^{\lambda A/2}+C_{3}\lambda^{3}+O(\lambda^{4}),\label{eq:splitting_formula_2}
\end{equation}
where $C_{3}$ is a combination of commutators of $A$ and $B$. A
direct fourth order splitting scheme was derived by Chin and Suzuki
\cite{chin1997splittingsymplectic,suzuki1995splittinghybrid}:

\begin{equation}
e^{\lambda(A+B)}=e^{\lambda\frac{1}{6}A}e^{\lambda\frac{1}{2}B}e^{\lambda\frac{2}{3}A+\lambda^{3}\frac{1}{72}[A,[B,A]]}e^{\lambda\frac{1}{2}B}e^{\lambda\frac{1}{6}A}+C_{5}\lambda^{5}+O(\lambda^{6}).\label{eq:splitting_formula_4}
\end{equation}
This splitting requires also the evaluation of the $[A,[B,A]]$ commutator,
which can rise additional difficulties, depending on the particular
form of $A$ and $B$.

We proceed by introducing another kind of higher order operator splitting,
based on the work of Bandrauk and Shen \cite{bandrauk1992splitting},
who developed an iterative method to improve the accuracy of the (\ref{eq:splitting_formula_2})
scheme. Let us denote the second order accurate form with $S_{2}(\lambda)=e^{\lambda A/2}e^{\lambda B}e^{\lambda A/2}$,
then their iteration method reads for $n=4,6,...$ as 
\begin{equation}
S_{n}(\lambda)=S_{n-2}(s\lambda)S_{n-2}((1-2s)\lambda)S_{n-2}(s\lambda)+C_{n-1}(2s^{n-1}+(1-2s)^{n-1})\lambda^{n-1}+O(\lambda^{n+1})\label{eq:splitting_iter_4}
\end{equation}
where the parameter $s$ must be for each iteration step a real root
of the corresponding polynomial equation 
\begin{eqnarray}
2s^{n-1}+(1-2s)^{n-1} & = & 0.\label{eq:splitting_iter_4_poly}
\end{eqnarray}
In (\ref{eq:splitting_iter_4}) only the odd error terms appear, because
of the unitarity and the symmetry of the splitting scheme. So the
$S_{n}(\lambda)$ requires $3^{n/2}$ evaluations of $S_{2}$, in
the worst case. For completeness we note, that using the complex roots
of \eqref{eq:splitting_iter_4_poly} in \eqref{eq:splitting_iter_4}
is also a viable approach in some numerical applications \cite{suzuki1995splittinghybrid,bandrauk2013splitting}.

This scheme was already generalized for time-dependent Hamiltonians
of the form $H(t)=A(t)+B(t)$ in \cite{bandrauk1992splitting,bandrauk1993splitting}
as follows. Inserting the second order effective Hamiltonian (\ref{eq:time_h_eff_2})
into the formula (\ref{eq:splitting_formula_2}) with $\lambda=-i\Delta t$
, the second order accurate splitting of the evolution operator becomes
\begin{equation}
U_{2}(t+\Delta t,\,\, t)=e^{-i\frac{\Delta t}{2}A(t+\Delta t/2)}e^{-i\,\Delta t\, B(t+\Delta t/2)}e^{-i\frac{\Delta t}{2}A(t+\Delta t/2)}.
\end{equation}
Then (\ref{eq:splitting_iter_4}) will take the form for $n=4,6...$
\begin{equation}
U_{n}(t+\Delta t,\,\, t)=U_{n-2}(t+\Delta t,\,\, t+(1-s)\Delta t)\,\, U_{n-2}(t+(1-s)\Delta t\,\,,t+s\Delta t)\,\, U_{n-2}(t+s\Delta t\,\,,t)+O(\lambda^{n+1})\label{eq:splitting_iter_4_u}
\end{equation}
and $s$ being the same as in the time-independent case (\ref{eq:splitting_iter_4_poly}).
Interestingly, the fourth order approximation $U_{4}$ in (\ref{eq:splitting_iter_4_u})
decreases the error even if the time evolution governed by a nonlinear
time-dependent Schrödinger equations, if the nonlinear error term
is corrected in the formulation of the $U_{2}$ propagator \cite{bandrauk1993splitting,bandrauk2013splitting}.

Now we write out a new form of iteration (\ref{eq:splitting_iter_4_u})
for time-dependent Hamiltonians, by using the same principles, for
$n=6,10,...$ 
\begin{align}
U_{n}(t+\Delta t,\,\, t)=\,\, & U_{n-4}(t+\Delta t,\,\, t+(1-s)\Delta t)\,\, U_{n-4}(t+(1-s)\Delta t,\,\, t+(1-s-p)\Delta t)\cdot\,\,\nonumber \\
\,\, & U_{n-4}(t+(1-s-p)\Delta t,\,\, t+(s+p)\Delta t)\cdot\,\,\nonumber \\
\,\, & U_{n-4}(t+(s+p)\Delta t,\,\, t+s\Delta t)\,\, U_{n-4}(t+s\Delta t,\,\, t)+O(\lambda^{n+1}),\label{eq:splitting_iter_6_u}
\end{align}
where $s$, $p$ must be the simultaneous real roots of equations
\begin{equation}
2s^{n-3}+2p^{n-3}+(1-2s-2p)^{n-3}=0,\,\,\,\,\,\text{and}\,\,\,\,\,\,2s^{n-1}+2p^{n-1}+(1-2s-2p)^{n-1}=0.
\end{equation}
However, this formula for $S_{6}$ requires five evaluations of $S_{2}$,
compared to nine in the case of the (\ref{eq:splitting_iter_4}).

Although the fourth order formula (\ref{eq:splitting_formula_4})
with the fourth order effective Hamiltonian (\ref{eq:time_h_eff_4})
seems to be superior compared to the iterative propagation (\ref{eq:splitting_iter_4_u})
(because of the extra information given by the temporal and spatial
commutators, and less evaluations), the schemes (\ref{eq:splitting_iter_4_u})
and (\ref{eq:splitting_iter_6_u}) do not require the calculation
of commutators, they decrease all the $\Delta t$ dependent errors
simultaneously and they are easy to implement. However, they involve
backward time steps, which means they do not work very well with diffusive
problems.

Another class of high order split-operator methods for diffusive partial
differential equations was developed in \cite{chin2009imagsplit},
where the exponential operator is found by an ansatz of 
\begin{equation}
S_{2n}(\lambda)=\sum_{k=1}^{n}c_{k}S_{2}^{k}\left(\frac{\lambda}{k}\right)+O\left(\lambda^{2n+1}\right)\text{ with }c_{k}=\prod_{j=1(\neq k)}^{n}\frac{c_{k}^{2}}{c_{k}^{2}-c_{j}^{2}}.\label{eq:splitting_imag_2n}
\end{equation}
Here $S_{2}$ is the same second order formula which is used in (\ref{eq:splitting_iter_4}).
The fourth order formula is simply given by $S_{4}(\lambda)=-\frac{1}{3}S_{2}(\lambda)+\frac{4}{3}S_{2}^{2}\left(\frac{\lambda}{2}\right)$.
Thus, once $S_{2}$ is properly constructed, all high order formulae
(\ref{eq:splitting_iter_4_u}), (\ref{eq:splitting_iter_6_u}), (\ref{eq:splitting_imag_2n})
can be utilized immediately. The (\ref{eq:splitting_imag_2n}) method
is suitable for the imaginary time propagation \cite{bauer2006tdseqprop,chin2009imagsplit,lehtovaara2007tdseimaginary}
with $\lambda\rightarrow-\Delta t$, i.e. for determination of the
lowest energy eigenstates of stationary potentials.

A problem also arises when one wishes to use an imaginary potential
\cite{muga2004absorbingpotentials} with (\ref{eq:splitting_iter_4_u})
and (\ref{eq:splitting_iter_6_u}), which is a commonly used numerical
technique to avoid artificial reflections from the (box) boundaries
of the simulation domain. Adding an imaginary potential $iV_{{\rm im}}$
into the Hamiltonian gives the following splitting of the exponential
operator from (\ref{eq:splitting_formula_2}): 
\begin{equation}
e^{-i\Delta t\, H+\Delta t\, V_{{\rm im}}}=e^{\Delta t\, V_{{\rm im}}/2}e^{-i\Delta t\, H}e^{\Delta t\, V_{{\rm im}}/2}+O(\Delta t^{3}).\label{eq:splitting_absorb_2}
\end{equation}
It is clear that $V_{{\rm im}}(z,\rho)<0$ performs the required absorption
of the wavefunction in the case of forward time propagation $\Delta t>0$,
but it would blow up the wave function during the necessary backward
steps in (\ref{eq:splitting_iter_4_u}) and (\ref{eq:splitting_iter_6_u}).
We circumvented this problem by partially disallowing backward steps,
i.e. we replaced $\Delta t\, V_{{\rm im}}$ by $|\Delta t|V_{{\rm im}}$.

\subsection{Directional splitting of the exponential operator\label{sub:splitting_direction_wise}}

The most common way of factorizing the exponential operator $e^{\lambda H}$
is that the different spatial coordinate derivatives decouple into
different exponential operators, i.e. to use a directional splitting.
Then the propagation can be carried out by solving multiple one dimensional
TDSE-s in succession. The most frequent realization of this is the
so-called potential-kinetic term splitting, which is mainly used in
 conjunction with the Fourier-transformation methods in Cartesian
coordinates \cite{leforestier1991tdsecomparison,blanes2000tdsesplitting,chin2001splittinggrad}.
However, if we apply the potential-kinetic term splitting to problem
of (\ref{eq:tdse_3dc}) then we have to write Hamiltonian as $H=T_{\rho}+T_{z}+V$
where $V$ is the potential, $T_{\rho}=\beta\partial_{\rho}^{2}+\beta\rho^{-1}\partial_{\rho}$
and $T_{z}=\beta\partial_{z}^{2}$ are the kinetic energy operators.
The $[A,[B,A]]$ commutator will take a rather simple form of $[V,[T_{\rho}+T_{z},V]]=-2\beta|\nabla V|^{2}$.
Thus, the direct second order (\ref{eq:splitting_formula_2}) and
fourth order (\ref{eq:splitting_formula_4}) symmetric splitting schemes
are written as 
\begin{equation}
e^{\lambda(T_{\rho}+T_{z}+V)}=e^{\lambda V/2}e^{\lambda T_{\rho}}e^{\lambda T_{z}}e^{\lambda V/2}+C_{3}\lambda^{3}+O(\lambda^{4}),\label{eq:splitting_direction_2}
\end{equation}

\begin{equation}
e^{\lambda(T_{\rho}+T_{z}+V)}=e^{\lambda\frac{1}{6}V}e^{\lambda\frac{1}{2}T_{\rho}}e^{\lambda\frac{1}{2}T_{z}}e^{\lambda\frac{2}{3}V+\lambda^{3}\frac{1}{72}\frac{1}{\mu}|\nabla V|^{2}}e^{\lambda\frac{1}{2}T_{\rho}}e^{\lambda\frac{1}{2}T_{z}}e^{\lambda\frac{1}{6}V}+C_{5}\lambda^{5}+O(\lambda^{6}),\label{eq:splitting_direction_4}
\end{equation}
where $\lambda=-i\Delta t$. We also note that these split-operator
formulae by themself will introduce error, scaling as $\Delta t^{3}$
and $\Delta t^{5}$, compared to the stationary states of the exact
time-independent Hamiltonian. In the case of a time-dependent Hamiltonian,
the respective second order (\ref{eq:time_h_eff_2}) or fourth order
(\ref{eq:time_h_eff_4}) effective Hamiltonians should be used. These
truncations of the Magnus-series (\ref{eq:time_h_magnus}) will introduce
additional $\Delta t^{3}$ and $\Delta t^{5}$ dependent errors in
time evolution.

The magnitudes of the aforementioned numerical errors will depend
on the derivatives of $V(z,\rho,t)$: in the case of an external electric
field the smoothness of $V$ in time is a reasonable assumption. On
the other hand, the spatial dependence of the atomic potential should
also behave the same way if we wish to use operator splitting. As
we suspect, it will not always be the case, especially in atomic or
molecular physics. We can deduce from (\ref{eq:splitting_direction_4})
that the leading order of error of the second order scheme must have
a dependence of $\Delta t^{3}|\nabla V|^{2}$, e.g. the error characteristics
strongly depend on the magnitude of the spatial derivatives of $V(z,\rho)$.
In the case of an atomic $1/r$ Coulomb potential this error term
will be $\Delta t^{3}r^{-4}$ which becomes significant only in the
region $r<1$, where it is increasing rapidly with fourth power of
$1/r$, and the split operator scheme completely breaks down at $r=0$.
This also illustrates the fact that non-differentiability of either
the potential or the wavefunction could potentially make operator
splitting like (\ref{eq:splitting_direction_2}), (\ref{eq:splitting_direction_4})
less-than-useful.

Up to now, we have seen that directional spitting related exponential
operator factorizations result drastic speed improvement for well
behaved potentials: we can directly apply the five-point finite difference
Crank-Nicolson method to the $e^{\lambda T_{z}}$ and $e^{\lambda T_{\rho}}$exponents
for evaluation, thus we have to solve for locally decoupled one dimensional
wave functions. Then, the approximate operations count of evaluating
the formula (\ref{eq:splitting_direction_2}) is $\sim N_{z}N_{\rho}$,
which is smaller by the factor of $N_{\rho}^{2}$ compared to the
full Crank-Nicolson problem of the same size. On the other hand, the
full Crank-Nicolson scheme is able to incorporate the Neumann and
Robin boundary conditions at $\rho=0$ into the implicit linear equations
properly, and it does not suffer from the catastrophic error blow
up while we approach the point Coulomb singularity.

In this article we propose a merger of a split-operator method with
the 2D Crank-Nicolson scheme, in the form of ``hybrid splitting'',
to get the best of both of these methods.

\subsection{Hybrid splitting of the exponential operator\label{sub:splitting_hybrid_scheme}}

\begin{figure}
\begin{centering}
\includegraphics[width=12cm]{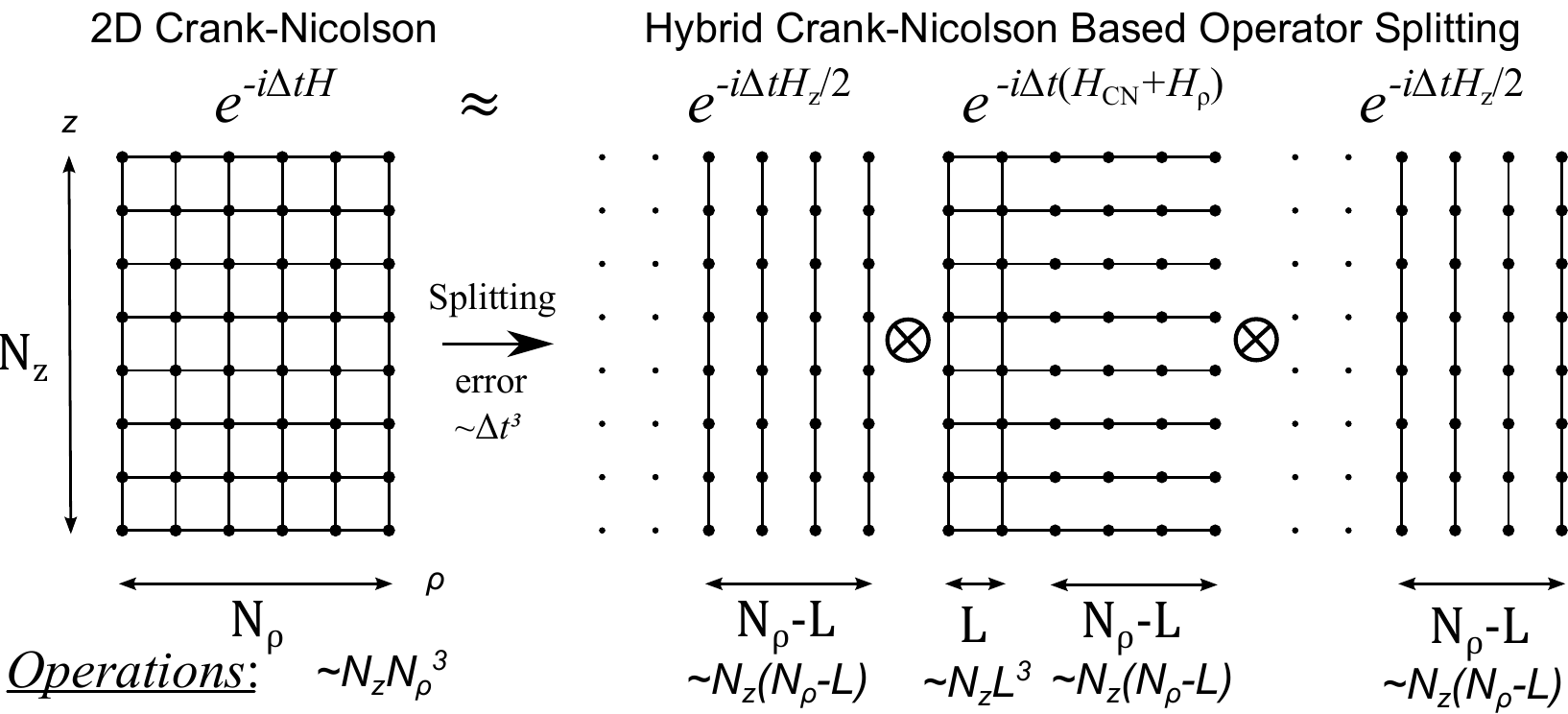} 
\par\end{centering}

\protect\caption{Sketch of our hybrid splitting scheme. The costly 2D Crank-Nicolson
scheme was replaced with a special second order symmetric split operator
formula except at the $L$ nearest gridpoints in the neighborhood
of the $\rho=0$ axis, in order to retain accuracy and stability.
The solid lines represent coupling between the gridpoints of the exponential
operator evaluations. We also indicate the approximate operations
count needed to solve the respective systems of linear equations.\label{fig:hybrid_splitting}}
\end{figure}

Let us split the spatial domain as $G=G_{\text{CN}}+G_{\text{Split}}$
where 
\begin{equation}
G=\{z,\rho\in\mathbb{R},\, z_{\min}\leq z\leq z_{\min}+N_{z}\Delta z,\,\,0\leq\rho\leq N_{\rho}\Delta\rho\},\label{eq:hybid_g}
\end{equation}
\begin{equation}
G_{\text{CN}}=\{z,\rho\in\mathbb{R},\, z_{\min}\leq z\leq z_{\min}+N_{z}\Delta z,\,\,0\leq\rho\leq L\Delta\rho\},\label{eq:hybrid_g_cn}
\end{equation}
\begin{equation}
G_{\text{\text{Split}}}=\{z,\rho\in\mathbb{R},\, z_{\min}\leq z\leq z_{\min}+N_{z}\Delta z,\,\, L\Delta\rho<\rho\leq N_{\rho}\Delta\rho\},\label{eq:hybrid_g_split}
\end{equation}
then we define the pieces of the Hamiltonian $H=H_{z}+H_{\rho}+H_{{\rm CN}}$
as

\begin{equation}
H_{z}=\beta\partial_{z}^{2}\text{ \,\,\ if }(z,\rho)\in G_{\text{Split}},\label{eq:hybrid_h_z}
\end{equation}
\begin{equation}
H_{\rho}=\beta\partial_{\rho}^{2}+\beta\rho^{-1}\partial_{\rho}+V(z,\rho,t)\text{ \,\,\ if }(z,\rho)\in G_{\text{Split}},\label{eq:hybrid_h_rho}
\end{equation}
\begin{equation}
H_{\text{CN}}=\beta\partial_{z}^{2}+\beta\partial_{\rho}^{2}+\beta\rho^{-1}\partial_{\rho}+V(z,\rho,t)\text{ \,\,\ if }(z,\rho)\in G_{\text{CN}}.\label{eq:hybrid_h_cn}
\end{equation}
Then the original Hamiltonian can be reconstructed as 
\begin{equation}
H=\begin{cases}
H_{z}+H_{\rho} & \text{if }(z,\rho)\in G_{\text{Split}}\\
H_{\text{CN}} & \text{if }(z,\rho)\in G_{\text{CN}}
\end{cases}.\label{eq:hybrid_h_full}
\end{equation}
$H$ will never get evaluated outside $G$, in accordance with the
boundary conditions for the wave function $\Psi(z,\rho,t)$.

Thus, we have partitioned the spatial domain into two regions: $G_{{\rm CN}}$
where (based on the previous section) we do not use any split operator
approximation and propagate with Hamiltonian $H_{{\rm CN}}$, and
region $G_{{\rm Split}}$ where we do use operator splitting as $H_{z}+H_{\rho}$.

In order to merge the directional operator splitting approach with
the true 2D Crank-Nicolson method, we introduce our second order hybrid
splitting scheme: 
\begin{equation}
e^{-i\Delta t(H_{z}+H_{\rho}+H_{\text{CN}})}=e^{-i\Delta t\, H_{z}/2}e^{-i\Delta t\,(H_{\rho}+H_{\text{CN}})}e^{-i\Delta t\, H_{z}/2}+O(\Delta t^{3}),\label{eq:hybrid_splitting_2}
\end{equation}
where we keep $H_{{\rm CN}}$ and $H_{\rho}$ in the same exponent,
in order that the wave function can ``freely flow'' between the
two regions $G_{{\rm CN}}$ and $G_{{\rm Split}}$ without introducing
further artifacts. (Note that the exponential operator $e^{-i\Delta t\,(H_{\rho}+H_{\text{CN}})}$
cannot be split further in this sense.) We use this scheme as the
second order terms in the iterative formulae (\ref{eq:splitting_iter_4}),
(\ref{eq:splitting_iter_6_u}), (\ref{eq:splitting_imag_2n}) to gain
higher order accuracy in $\Delta t$. The $e^{-i\Delta t\, H_{z}/2}$
part can be evaluated with any method of choice. We have constructed
a Numerov-extended Crank-Nicolson line propagation algorithm for this,
which can be found in \ref{sub:numerov_z_line}.

In order to evaluate the $e^{-i\Delta t\,(H_{\rho}+H_{\text{CN}})}$
operator, we apply the second order Padé-approximation (\ref{eq:time_pade_cn_2})
to arrive again at a second order Crank-Nicolson form of 
\begin{equation}
\Psi(t_{k+1})=e^{-i\Delta t\,(H_{\rho}+H_{\text{CN}})}\Psi(t_{k})=\left(1+\alpha(H_{\rho}+H_{\text{CN}})\right)^{-1}\left(1-\alpha(H_{\rho}+H_{\text{CN}})\right)\Psi(t_{k})+O(\Delta t^{3})\label{eq:hybrid_pade_2}
\end{equation}
where $\alpha=i\Delta t/2$ and $V(z,\rho,t)$ is evaluated at the
midpoint $t+\Delta t/2$. We introduce the spatial grids (\ref{eq:cn_spatial_z}),
(\ref{eq:cn_spatial_r}) and the discrete operators $L_{z}$, $L_{\rho}$
and $D_{\rho}$, then we get the following equations for the two regions
for $i\in[0,N_{z}]$: 
\begin{equation}
\left(1+\alpha\beta L_{\rho}+\alpha\beta L_{z}+\alpha V_{i,j}\right)\Psi_{i,j}(t_{k+1})=\left(1-\alpha\beta L_{\rho}-\alpha\beta L_{z}-\alpha V_{i,j}\right)\Psi_{i,j}(t_{k}),\mbox{\text{ if }}j\in[1,L],\label{eq:hybrid_cn_pade_2}
\end{equation}
\begin{equation}
\left(1+\alpha\beta L_{\rho}+V_{i,j}\right)\Psi_{i,j}(t_{k+1})=\left(1-\alpha\beta L_{\rho}-\alpha V_{i,j}\right)\Psi_{i,j}(t_{k}),\mbox{\text{ if }}j\in[L+1,N_{\rho}].\label{eq:hybrid_rho_pade_2}
\end{equation}
Using five point finite differences, the first set of equations can
be expanded resulting in the form of (\ref{eq:cn_3dc_full}), and
the expansion of the second set gives the following similar result:

\[
(-1+1/j)\beta_{\rho}X_{i,j-2}+(16-8/j)\beta_{\rho}X_{i,j-1}+(1-30\beta_{\rho}+\alpha V_{i,j})X_{i,j}+
\]
\[
(16+8/j)\beta_{\rho}X_{i,j+1}+(-1-1/j)\beta_{\rho}X_{i,j+2}=(1-1/j)\beta_{\rho}\Psi_{i,j-2}+(-16+8/j)\beta_{\rho}\Psi_{i,j-1}+
\]
\begin{equation}
(1+30\beta_{\rho}-\alpha V_{i,j})\Psi_{i,j}+(-16-8/j)\beta_{\rho}\Psi_{i,j+1}+(1+1/j)\beta_{\rho}\Psi_{i,j+2}.\label{eq:cn_rho_full}
\end{equation}
With the presence of a Coulomb-core at $z_{R}=0$, the boundary condition
at $\rho=0$ will be once again for the $m=0$ configuration 
\begin{equation}
\left(D_{\rho}+\gamma/(2\beta)\cdot\delta_{R,i}\right)\Psi_{i,0}(t_{k})=0\label{eq:hybrid_bc_nm}
\end{equation}
with the same expanded form as (\ref{eq:cn_3dc_full_nm_cb}). For
the $m\neq0$ states, we just use the Dirichlet-boundary condition
(\ref{eq:cn_3dc_full_bc_m}) instead of (\ref{eq:hybrid_bc_nm}).
The box boundary conditions also apply for every $m$: 
\begin{equation}
\Psi_{i,j}(t_{k+1})=0,\text{ if }i\notin[0,N_{z}],j\notin[-N_{\rho},N_{\rho}].\label{eq:hybrid_bc_box}
\end{equation}
 From these, a mixed 2D - 1D Crank-Nicolson scheme can be constructed
in the $G_{{\rm CN}}+G_{{\rm Split}}$ domain depending on $m$, which
is fourth order accurate in space and second order accurate in time.

What are the advantages of this scheme? First, if $L>1/\Delta\rho$
the accuracy of the directional splitting can be considerably increased
in the presence of the Coulomb potential, because we removed directional
splitting near its core (where its gradient is the largest). Second,
if $L\geq5$ (the ``width'' of $D_{\rho}$) then the (\ref{eq:hybrid_bc_nm})
condition no longer affects into the split operator zone: we will
maintain stability. Third, the operation-count is approximately $\sim L^{3}N_{z}$
plus $\sim(N_{\rho}-L)N_{z}$, meaning if $L\ll N_{\rho}$ then we
can regain the speed of the directional splitting and large part of
the algorithm, corresponding to equations (\ref{eq:hybrid_rho_pade_2})
can be parallelized for different $j$ indices. So, if we set $L$
to the smallest value sufficient for accuracy then we can acquire
a very efficient scheme.

However, it is not straightforward to solve these linear equations
effectively, therefore we present as special algorithm for this, which
we call ``hybrid splitting solver algorithm''.

\section{Hybrid splitting solver algorithm\label{sec:hybrid_splitting_algorithm}}

In this Section, we write out the matrix form of the linear equations
resulting from the approximation of the exponential operator $e^{-i\Delta t\,(H_{\rho}+H_{\text{CN}})}$
to familiarize ourselves with its structure, which is required for
developing an efficient propagation algorithm for our hybrid splitting
scheme. Then we outline the solution algorithm.

\subsection{The matrix form of the linear equations\label{sub:hs_algorithm_equations_full}}

Recalling the notations $X_{i,j}=\Psi_{i,j}(t+\Delta t)$, $\Psi_{i,j}=\Psi_{i,j}(t)$,
let us construct the column vectors corresponding to the $i$th row
of the 2D problem as 
\begin{equation}
\boldsymbol{\Psi}_{i}=\begin{pmatrix}\Psi_{i,0} & \Psi_{i,1} & \Psi_{i,2} & \ldots & \Psi_{i,N_{\rho}}\end{pmatrix}^{\mathrm{T}}\text{ and }\boldsymbol{X}_{i}=\begin{pmatrix}X_{i,0} & X_{i,1} & X_{i,2} & \ldots & X_{i,N_{\rho}}\end{pmatrix}^{\mathrm{T}}.\label{eq:hs_alg_full_vec_psi_x}
\end{equation}
Then, the joint problem (\ref{eq:hybrid_cn_pade_2}), (\ref{eq:hybrid_rho_pade_2})
will take the block pentadiagonal form of 
\begin{equation}
\begin{bmatrix}{\rm B}_{0} & {\rm C_{0}} & {\rm F_{0}} & 0 & 0 & \ldots & 0\\
{\rm A}_{1} & {\rm B}_{1} & {\rm C_{1}} & {\rm F_{1}} & 0 & \ldots & 0\\
{\rm E_{2}} & {\rm A}_{2} & {\rm B}_{2} & {\rm C_{2}} & {\rm F_{2}} & \ldots & 0\\
\vdots & \ddots & \ddots & \ddots & \ddots & \ddots & \vdots\\
0 & \ldots & {\rm E}_{N_{z}-2} & {\rm A}_{N_{z}-2} & {\rm B}_{N_{z}-2} & {\rm {\rm C}}_{N_{z}-2} & {\rm {\rm F}}_{N_{z}-2}\\
0 & \ldots & 0 & {\rm E}_{N_{z}-1} & {\rm A}_{N_{z}-1} & {\rm B}_{N_{z}-1} & {\rm C}_{N_{z}-1}\\
0 & \ldots & 0 & 0 & {\rm E}_{N_{z}} & {\rm A}_{N_{z}} & {\rm B}_{N_{z}}
\end{bmatrix}\begin{bmatrix}\boldsymbol{X}_{0}\\
\boldsymbol{X}_{1}\\
\boldsymbol{X}_{2}\\
\vdots\\
\boldsymbol{X}_{N_{z}-2}\\
\boldsymbol{X}_{N_{z}-1}\\
\boldsymbol{X}_{N_{z}}
\end{bmatrix}=\begin{bmatrix}{\bf y}_{0}\\
{\bf y}_{1}\\
{\bf y}_{2}\\
\vdots\\
{\bf y}_{N_{z}-2}\\
{\bf y}_{N_{z}-1}\\
{\bf y}_{N_{z}}
\end{bmatrix}.\label{eq:hs_alg_matrix_big}
\end{equation}
Here ${\rm E}_{i},{\rm A}_{i},{\rm B}_{i},{\rm C_{i}},{\rm F_{i}}$
are $(N_{\rho}+1)\times(N_{\rho}+1)$ matrices. Particularly, 
\begin{equation}
{\rm E_{i}=}\begin{cases}
\text{diag}(0,e_{z,i,1},\dots,e_{z,i,L},0,\dots,0) & \text{ if }i\in[2,N_{z}],\end{cases}\label{eq:hs_alg_matrix_big_e}
\end{equation}
\begin{equation}
{\rm C_{i}=\begin{cases}
\text{diag}(0,c_{z,i,1},\dots,c_{z,i,L},0,\dots,0) & \text{ if }i\in[1,N_{z}],\end{cases}}\label{eq:hs_alg_matrix_big_c}
\end{equation}
\begin{equation}
{\rm A_{i}=\begin{cases}
\text{diag}(0,a_{z,i,1},\dots,a_{z,i,L},0,\dots,0) & \text{ if }i\in[0,N_{z}-1],\end{cases}}\label{eq:hs_alg_matrix_big_a}
\end{equation}
\begin{equation}
{\rm F_{i}=\begin{cases}
\text{diag}(0,f_{z,i,1},\dots,f_{z,i,L},0,\dots,0) & \text{ if }i\in[0,N_{z}-2],\end{cases}}\label{eq:hs_alg_matrix_big_f}
\end{equation}
are the diagonal matrices responsible for coupling the adjacent $\rho$-rows
(with different values of coordinate $z$), and ${\rm B}_{i}$ is
an almost five-diagonal matrix in the form of

\begin{equation}
{\rm B}_{i}=\begin{bmatrix}d_{0,i} & d_{1,i} & d_{2,i} & d_{3,i} & d_{4,i} & \ldots & 0\\
a_{i,0} & b_{i,1} & c_{i,1} & f_{i,1} & 0 & \ldots & 0\\
e_{i,2} & a_{i,2} & b_{i,2} & c_{i,2} & f_{i,2} & \ldots & 0\\
\vdots & \ddots & \ddots & \ddots & \ddots & \ddots & \vdots\\
0 & \ldots & e_{i,N_{\rho}-2} & a_{i,N_{\rho}-2} & b_{i,N_{\rho}-2} & c_{i,N_{\rho}-2} & f_{i,N_{\rho}-2}\\
0 & \ldots & 0 & e_{i,N_{\rho}-1} & a_{i,N_{\rho}-1} & b_{i,N_{\rho}-1} & c_{i,N_{\rho}-1}\\
0 & \ldots & 0 & 0 & e_{i,N_{\rho}} & a_{i,N_{\rho}} & b_{i,N_{\rho}}
\end{bmatrix}.\label{eq:hs_alg_matrix_big_b}
\end{equation}
In the above matrices we have already taken into account the box,
the Neumann and Robin boundary conditions for the $m=0$ configuration.
The $d_{0,i},d_{1,i},d_{2,i},d_{3,i},d_{4,i}$ coefficients are given
by the expanded equation (\ref{eq:cn_3dc_full_nm_cb}) as

\begin{equation}
d_{0,i}=-25+6(\gamma/\beta)\Delta\rho\cdot\delta_{R,i},\,\,\, d_{1,i}=48,\,\,\, d_{2,i}=-36,\,\,\, d_{3,i}=16,\,\,\, d_{4,i}=-3\label{eq:hs_alg_coeff_d}
\end{equation}
for all $i\in[0,N_{z}]$. These coefficients take the diagonal form
of $d_{j,i}=\delta_{j,0}$ for the $m\neq0$ states.

In the Crank-Nicolson region ($j\leq L$), the coefficients are given
by equation (\ref{eq:cn_3dc_full}):

\begin{equation}
b_{i,j}=1-30\beta_{\rho}-30\beta_{z}+\alpha V_{i,j},\,\,\, a_{z,i,j}=c_{z,i,j}=16\beta_{z}\,\,\,\, e_{z,i,j}=f_{z,i,j}=-\beta_{z},\label{eq:hs_alg_coeff_b_cn}
\end{equation}
\begin{equation}
e_{i,j}=(-1+1/j)\beta_{\rho},\,\,\, f_{i,j}=(-1-1/j)\beta_{\rho},\label{eq:hs_alg_coeff_ef}
\end{equation}
\begin{equation}
a_{i,j}=(16-8/j)\beta_{\rho},\,\,\, c_{i,j}=(16+8/j)\beta_{\rho}.\label{eq:hs_alg_coeff_ac}
\end{equation}
In the split region ($j\geq L+1$), we inspect the equation (\ref{eq:cn_rho_full}),
and note that only $b_{i,j}$ get modified as 
\begin{equation}
b_{i,j}=1-30\beta_{\rho}+\alpha V_{i,j}\text{\,\,\,\,\ if\,\,\,}i\in[L+1,N_{\rho}],j\in[0,N_{z}]\label{eq:hs_alg_coeff_b_rho}
\end{equation}
and the $a_{i,j}$, $c_{i,j}$, $e_{i,j}$, $f_{i,j}$ coefficients
are the same as in (\ref{eq:hs_alg_coeff_ef}), (\ref{eq:hs_alg_coeff_ac}).

The right hand sides are given by (\ref{eq:cn_3dc_full}) and (\ref{eq:cn_rho_full}),
which can be written in a somewhat simpler matrix form of 
\begin{equation}
{\bf y}_{i}=2\boldsymbol{\Psi}_{i}-{\rm E}_{i}\boldsymbol{\Psi}_{i-2}-{\rm A}_{i}\boldsymbol{\Psi}_{i-1}-{\rm B}_{i}\boldsymbol{\Psi}_{i}-{\rm C}_{i}\boldsymbol{\Psi}_{i+1}-{\rm F}_{i}\boldsymbol{\Psi}_{i+2}\text{ with }y_{i,0}=0,\label{eq:hs_alg_vector_rhs}
\end{equation}
which also takes into account the boundary conditions.

\subsection{Reducing the number of the equations}

One can see that the directional splitting introduced $N_{\rho}-L$
zeros at the end of the diagonal of the matrices ${\rm E}_{i},{\rm A}_{i},{\rm C_{i}},{\rm F_{i}}$
which means that the corresponding $\rho$-lines are not directly
coupled. Taking advantage of this we significantly increase the computational
efficiency by eliminating the improper matrix elements to reduce the
effective block size to $(L+1)^{2}$.

To proceed, we take the equations corresponding to the $i$th block
matrix row 
\begin{equation}
\left[\begin{array}{ccccccc}
\ldots & {\rm E}_{i} & {\rm A}_{i} & {\rm B}_{i} & {\rm C}_{i} & {\rm F}_{i} & \ldots\end{array}\right]\cdot{\bf X}={\bf y}_{i}\label{eq:hs_alg_elim_block_row}
\end{equation}
and write out their coefficient matrix from rows $L-1$ to $N_{\rho}$:

\begin{equation}
\left[\begin{array}{ccccccccccc}
\ldots & e_{i,L-1} & a_{i,L-1} & b_{i,L-1} & c_{i,L-1} & f_{i,L-1} & 0 & 0 & \ldots & 0 & \ldots\\
\ldots & 0 & e_{i,L} & a_{i,L} & b_{i,L} & c_{i,L} & f_{i,L} & 0 & \ldots & 0 & \ldots\\
\ldots & 0 & 0 & e_{i,L+1} & a_{i,L+1} & b_{i,L+1} & c_{i,L+1} & f_{i,L+1} & \ldots & 0 & \ldots\\
\ldots & \vdots & \vdots & \vdots & \ddots & \ddots & \ddots & \ddots & \ddots & \vdots & \ldots\\
\ldots & 0 & 0 & 0 & \ldots & e_{i,N_{\rho}-2} & a_{i,N_{\rho}-2} & b_{i,N_{\rho}-2} & c_{i,N_{\rho}-2} & f_{i,N_{\rho}-2} & \ldots\\
\ldots & 0 & 0 & 0 & \ldots & 0 & e_{i,N_{\rho}-1} & a_{i,N_{\rho}-1} & b_{i,N_{\rho}-1} & c_{i,N_{\rho}-1} & \ldots\\
\ldots & 0 & 0 & 0 & \ldots & 0 & 0 & e_{i,N_{\rho}} & a_{i,N_{\rho}} & b_{i,N_{\rho}} & \ldots
\end{array}\right].\label{eq:hs_alg_elim_matrix}
\end{equation}
Here we remind that rows $j\leq L$ have extra nonzero entries far
away from the diagonal, cf. (\ref{eq:hs_alg_matrix_big_e})-(\ref{eq:hs_alg_matrix_big_f}).
These lines cannot be used during the row operations, but the rest
of them, with $j>L$ can be used. To reduce Eq. (\ref{eq:hs_alg_matrix_big})
to a smaller block five-diagonal problem, $f_{i,L-1}$, $c_{i,L}$,
$f_{i,L}$ must be eliminated for all $i\in[0,N_{z}]$. Then, the
solution in the 2D Crank-Nicolson region $G_{{\rm CN}}$ will no longer
depend on the solution in the directionally split region $G_{{\rm Split}}$.
The structure of the ${\rm B}_{i}$ matrix makes it possible to use
the following backward elimination process from the $N_{\rho}$th
equation on, for all $i\in[0,N_{z}]$:

\begin{equation}
\left[\begin{array}{ccccccccccc}
\ldots & e_{i,L-1} & a_{i,L-1} & \tilde{b}_{i,L-1} & \tilde{c}_{i,L-1} & 0 & \ldots & 0 & 0 & 0 & \ldots\\
\ldots & 0 & e_{i,L} & \tilde{a}_{i,L} & \tilde{b}_{i,L} & 0 & \ldots & 0 & 0 & 0 & \ldots\\
\ldots & 0 & 0 & e_{i,L+1} & \tilde{a}_{i,L+1} & \tilde{b}_{i,L+1} & \ldots & 0 & 0 & 0 & \ldots\\
\ldots & \vdots & \vdots & \vdots & \ddots & \ddots & \ddots & \ddots & \ddots & \vdots & \ldots\\
\ldots & 0 & 0 & 0 & \ldots & e_{i,N_{\rho}-2} & \tilde{a}_{i,N_{\rho}-2} & \tilde{b}_{i,N_{\rho}-2} & 0 & 0 & \ldots\\
\ldots & 0 & 0 & 0 & \ldots & 0 & e_{i,N_{\rho}-1} & \tilde{a}_{i,N_{\rho}-1} & \tilde{b}_{i,N_{\rho}-1} & 0 & \ldots\\
\ldots & 0 & 0 & 0 & \ldots & 0 & 0 & e_{i,N_{\rho}} & \tilde{a}_{i,N_{\rho}} & \tilde{b}_{i,N_{\rho}} & \ldots
\end{array}\right]\label{eq:hs_alg_elim_matrix_lower}
\end{equation}
with right hand side of 
\begin{equation}
\widetilde{\boldsymbol{y}}_{i}=\begin{pmatrix}y_{i,0} & \dots & y_{i,L-2} & \tilde{y}_{i,L-1} & \tilde{y}_{i,L} & \tilde{y}_{i,L+1} & \ldots & \tilde{y}_{i,N_{\rho}-1} & \tilde{y}_{i,N_{\rho}}\end{pmatrix}^{\mathrm{T}},\label{eq:hs_alg_elim_rhs}
\end{equation}
where 
\begin{equation}
\tilde{c}_{i,j}=\begin{cases}
c_{i,j} & \text{ if }j=N_{\rho}-1\\
c_{i,j}-(f_{i,j}/\tilde{b}_{i,j+2})\tilde{a}_{i,j+2} & \text{ if }j=N_{\rho}-2\dots L-1,
\end{cases}\label{eq:hs_alg_elim_c}
\end{equation}

\begin{equation}
\tilde{a}_{i,j}=\begin{cases}
a_{i,j} & \text{ if }j=N_{\rho}\\
a_{i,j}-(\tilde{c}_{i,j}/\tilde{b}_{i,j+1})e_{i,j+1} & \text{ if }j=N_{\rho}-1\dots L,
\end{cases}\label{eq:hs_alg_elim_a}
\end{equation}
\begin{equation}
\tilde{b}_{i,j}=\begin{cases}
b_{i,j} & \text{ if }j=N_{\rho}\\
b_{i,j}-(\tilde{c}_{i,j}/\tilde{b}_{i,j+1})\tilde{a}_{i,j+1} & \text{ if }j=N_{\rho}-1\\
b_{i,j}-(\tilde{c}_{i,j}/\tilde{b}_{i,j+1})\tilde{a}_{i,j+1}-(f_{i,j}/\tilde{b}_{i,j+2})e_{i,j+2} & \text{ if }j=N_{\rho}-2\dots L\\
b_{i,j}-(f_{i,j}/\tilde{b}_{i,j+2})e_{i,j+2} & \text{ if }j=L-1,
\end{cases}\label{eq:hs_alg_elim_b}
\end{equation}
\begin{equation}
\tilde{y}_{i,j}=\begin{cases}
y_{i,j} & \text{ if }j=N_{\rho}\\
y_{i,j}-(\tilde{c}_{i,j}/\tilde{b}_{i,j+1})\tilde{y}_{i,j+1} & \text{ if }j=N_{\rho}-1\\
y_{i,j}-(\tilde{c}_{i,j}/\tilde{b}_{i,j+1})\tilde{y}_{i,j+1}-(f_{i,j}/\tilde{b}_{i,j+2})\tilde{y}_{i,j+2} & \text{ if }j=N_{\rho}-2\dots L\\
y_{i,j}-(f_{i,j}/\tilde{b}_{i,j+2})\tilde{y}_{i,j+2} & \text{ if }j=L-1,
\end{cases}\label{eq:hs_alg_elim_y}
\end{equation}
and the rest of the values remain unchanged. During the process, $\tilde{c}_{i,j}$
needs to be calculated first, then $\tilde{a}_{i,j}$, $\tilde{b}_{i,j},$
$\tilde{y}_{i,j}$ can be evaluated. Additionally, the value of $\tilde{c}_{i,j}$
is needed only in one step, then it can be discarded.

When the reduced equations are solved (cf. the next Section), we can
solve for all variables with forward substitution: 
\begin{equation}
X_{i,j}=\begin{cases}
\text{solution of the reduced block five diagonal part} & \text{ if }j\leq L\\
\left(\tilde{y}_{i,j}-\tilde{a}_{i,j}X_{i,j-1}-e_{i,j}X_{i,j-2}\right)/\tilde{b}_{i,j} & \text{ if }j=L+1\dots N_{\rho}
\end{cases}.\label{eq:hs_alg_elim_solution}
\end{equation}
These formulae above can be obtained by disassembling existing five-diagonal
solvers (for example \cite{karawia2006solution5diagonal}), however,
care must be taken handling the boundary values at the $j=L$ edge
and in the forward substitution afterwards.

Because of the special structure of the coefficient matrix in (\ref{eq:hs_alg_matrix_big}),
this process of backward elimination and forward substitution can
be parallelized to $N_{z}+1$ independent threads. They are depending
on a synchronization step though, which consists of the solution of
the following block five-diagonal part.

\subsection{The reduced system to solve}

After we performed the elimination procedure for every block ${\rm B}_{i}$,
we obtain a block five-diagonal system just like (\ref{eq:hs_alg_matrix_big})
with a drastically reduced block size, in the following form:

\begin{equation}
\begin{bmatrix}{\rm \widetilde{B}}_{0} & {\rm \widetilde{C}_{0}} & {\rm \widetilde{F}_{0}} & 0 & 0 & \ldots & 0\\
\widetilde{{\rm A}}_{1} & {\rm \widetilde{B}}_{1} & {\rm \widetilde{C}_{1}} & {\rm \widetilde{F}_{1}} & 0 & \ldots & 0\\
{\rm \widetilde{E}_{2}} & \widetilde{{\rm A}}_{2} & {\rm \widetilde{B}}_{2} & {\rm \widetilde{C}_{2}} & {\rm \widetilde{F}_{2}} & \ldots & 0\\
\vdots & \ddots & \ddots & \ddots & \ddots & \ddots & \vdots\\
0 & \ldots & {\rm \widetilde{E}}_{N_{z}-2} & \widetilde{{\rm A}}_{N_{z}-2} & {\rm \widetilde{B}}_{N_{z}-2} & {\rm \widetilde{C}}_{N_{z}-2} & {\rm \widetilde{F}}_{N_{z}-2}\\
0 & \ldots & 0 & {\rm \widetilde{E}}_{N_{z}-1} & \widetilde{{\rm A}}_{N_{z}-1} & {\rm \widetilde{B}}_{N_{z}-1} & \widetilde{C}_{N_{z}-1}\\
0 & \ldots & 0 & 0 & {\rm \widetilde{E}}_{N_{z}} & \widetilde{{\rm A}}_{N_{z}} & {\rm \widetilde{B}}_{N_{z}}
\end{bmatrix}\begin{bmatrix}\widetilde{\boldsymbol{X}}_{0}\\
\widetilde{\boldsymbol{X}}_{1}\\
\widetilde{\boldsymbol{X}}_{2}\\
\vdots\\
\widetilde{\boldsymbol{X}}_{N_{z}-2}\\
\widetilde{\boldsymbol{X}}_{N_{z}-1}\\
\widetilde{\boldsymbol{X}}_{N_{z}}
\end{bmatrix}=\begin{bmatrix}\widetilde{{\bf y}}_{0}\\
\widetilde{{\bf y}}_{1}\\
\widetilde{{\bf y}}_{2}\\
\vdots\\
\widetilde{{\bf y}}_{N_{z}-2}\\
\widetilde{{\bf y}}_{N_{z}-1}\\
\widetilde{{\bf y}}_{N_{z}}
\end{bmatrix}.\label{eq:hs_alg_matrix_small}
\end{equation}
These block matrices read as:

\begin{equation}
{\rm \widetilde{E}_{i}=}\begin{cases}
\text{diag}(0,e_{z,i,1},\dots,e_{z,i,L}) & \text{ if }i\in[2,N_{z}],\end{cases}\label{eq:hs_alg_small_e_block}
\end{equation}
\begin{equation}
{\rm \widetilde{C}_{i}}=\begin{cases}
\text{diag}(0,c_{z,i,1},\dots,c_{z,i,L}) & \text{ if }i\in[1,N_{z}],\end{cases}\label{eq:hs_alg_small_c_block}
\end{equation}
\begin{equation}
{\rm \widetilde{A}_{i}}=\begin{cases}
\text{diag}(0,a_{z,i,1},\dots,a_{z,i,L}) & \text{ if }i\in[0,N_{z}-1],\end{cases}\label{eq:hs_alg_small_a_block}
\end{equation}
\begin{equation}
{\rm \widetilde{F}_{i}}=\begin{cases}
\text{diag}(0,f_{z,i,1},\dots,f_{z,i,L}) & \text{ if }i\in[0,N_{z}-2],\end{cases}\label{eq:hs_alg_small_f_block}
\end{equation}
\begin{equation}
\widetilde{{\rm B}}_{i}=\begin{bmatrix}d_{0,i} & d_{1,i} & d_{2,i} & d_{3,i} & d_{4,i} & \ldots & 0\\
a_{i,0} & b_{i,1} & c_{i,1} & f_{i,1} & 0 & \ldots & 0\\
e_{i,2} & a_{i,2} & b_{i,2} & c_{i,2} & f_{i,2} & \ldots & 0\\
\vdots & \ddots & \ddots & \ddots & \ddots & \ddots & \vdots\\
0 & \ldots & e_{i,L-2} & a_{i,L-2} & b_{i,L-2} & c_{i,L-2} & f_{i,L-2}\\
0 & \ldots & 0 & e_{i,L-1} & a_{i,L-1} & b_{i,L-1} & \tilde{c}_{i,L-1}\\
0 & \ldots & 0 & 0 & e_{i,L} & \tilde{a}_{i,L} & \tilde{b}_{i,L}
\end{bmatrix},\label{eq:hs_alg_small_b_block}
\end{equation}
\begin{equation}
{\bf \widetilde{y}}_{i}=\begin{pmatrix}y_{i,0} & y_{i,1} & \ldots & y_{i,L-2} & \tilde{y}_{i,L-1} & \tilde{y}_{i,L}\end{pmatrix}^{\mathrm{T}},\label{eq:hs_alg_small_vec_y}
\end{equation}
\begin{equation}
\widetilde{\boldsymbol{X}}_{i}=\begin{pmatrix}X_{i,0} & X_{i,2} & \ldots & X_{i,L-2} & X_{i,L-1} & X_{i,L}\end{pmatrix}^{\mathrm{T}}.\label{eq:hs_alg_small_vec_x}
\end{equation}

The method of acquiring the solution to this system can be chosen
freely. For example, it can be solved by applying Gaussian elimination
(with forward substitution) to the $(L+1)^{2}$ matrix blocks or directly
to the $2L+1$ wide diagonal matrix. In either case, the time to obtain
the solution is drastically reduced (if $L\ll N_{\rho}$). After acquiring
the solution of (\ref{eq:hs_alg_matrix_small}) we complete the hybrid
splitting solver algorithm by formula (\ref{eq:hs_alg_elim_solution}),
as indicated in the previous Section.

\section{Numerical results\label{sec:numerical_results_main}}

Now let us turn our attention to the numerical test results of the
hybrid splitting method. First, in Section \ref{sub:test_3d_energies},
we investigate the errors related to the spatial discretization. Section
\ref{sub:test_3d_harmonic} is devoted to the tests of the time propagation
errors. In Section \ref{sub:test_3d_coulomb} we demonstrate the performance
of our algorithm through a real-world example.

\subsection{Stationary state energies\label{sub:test_3d_energies}}

To investigate the errors related to the spatial discretization, we
numerically compute the stationary states of selected test potentials.
We calculate energy errors caused by the finite differences by comparing
the numerical and the exact energy eigenvalues for certain energy
eigenstates. We have also set $\Delta\rho=\Delta z$.

\begin{table}[h]
\begin{centering}
\begin{tabular*}{1\textwidth}{@{\extracolsep{\fill}}ccccccc}
\hline 
\multicolumn{7}{c}{Target stationary states}\tabularnewline
\hline 
ID & Potential & Quantum Numbers & Energy & State & Normalization & Parameters\tabularnewline
\hline 
C-1s & $-\gamma/r$ & $n=1,l=0,m=0$ & $-\gamma^{2}\mu/2$ & $e^{-\mu\gamma r}$ & $\pi^{-1/2}(\mu\gamma)^{3/2}$ & $\mu=1,\gamma=1$\tabularnewline
C-2s & $-\gamma/r$ & $n=2,l=0,m=0$ & $-\gamma^{2}\mu/4$ & $[2-\mu\gamma r]e^{-\mu\gamma r/2}$ & $2^{-5/2}\pi^{-1/2}(\mu\gamma)^{3/2}$ & $\mu=1,\gamma=1$\tabularnewline
~~C-${\rm 2p}_{z}$ & $-\gamma/r$ & $n=2,l=1,m=0$ & $-\gamma^{2}\mu/4$ & $ze^{-\mu\gamma r/2}$ & $2^{-5/2}\pi^{-1/2}(\mu\gamma)^{5/2}$ & $\mu=1,\gamma=1$\tabularnewline
~~C-${\rm 2p}_{x}$ & $-\gamma/r$ & $n=2,l=1,m=1$ & $-\gamma^{2}\mu/4$ & $\rho e^{-\mu\gamma r/2}\cos\phi$ & $2^{-5/2}\pi^{-1/2}(\mu\gamma)^{5/2}$ & $\mu=1,\gamma=1$\tabularnewline
H-000 & $\frac{1}{2}\mu\omega^{2}r^{2}$ & $n_{x}=n_{y}=0,n_{z}=0$ & $\frac{3}{2}\omega$ & $e^{-\mu\omega r^{2}/2}$ & $\pi^{-3/4}(\mu\omega)^{3/4}$ & $\mu=1,\omega=1$\tabularnewline
H-001 & $\frac{1}{2}\mu\omega^{2}r^{2}$ & $n_{x}=n_{y}=0,n_{z}=1$ & $\frac{5}{2}\omega$ & $ze^{-\mu\omega r^{2}/2}$ & $2^{1/2}\pi^{-3/4}(\mu\omega)^{5/4}$ & $\mu=1,\omega=1$\tabularnewline
\hline 
\end{tabular*}
\par\end{centering}

\protect\caption{Properties of the potentials and their bound states that we used for
testing the errors regarding the spatial discretization. Our main
test potentials are the Coulomb potential (C) and the 3D harmonic
oscillator potential (H).\label{tab:test_3d_energies_cases}}
\end{table}

In these tests we have used our implementation of the true singular
Coulomb potential (C). Another test case, called ``Soft-Coulomb''
potential (SC), differs from C only in that the boundary condition
(\ref{eq:hybrid_bc_nm}) at the origin is replaced by (\ref{eq:cn_3dc_bc_nm})
at the origin. The results of both the C and the SC test cases are
compared to the exact eigenenergies and eigenstates of $1s$, $2s$,
$2p_{z}$. We also compute the eigenenergy of the state $2p_{x}$,
where C and SC are identical (see Section \ref{sec:tdse_3d_state_m}).
The energy errors of the 3D quantum harmonic oscillator (H) are also
tested for the ground and a first exited state. The most important
features of our test potentials are summarized in Table \ref{tab:test_3d_energies_table}.

\begin{figure}[h]
\begin{centering}
\includegraphics[width=13cm]{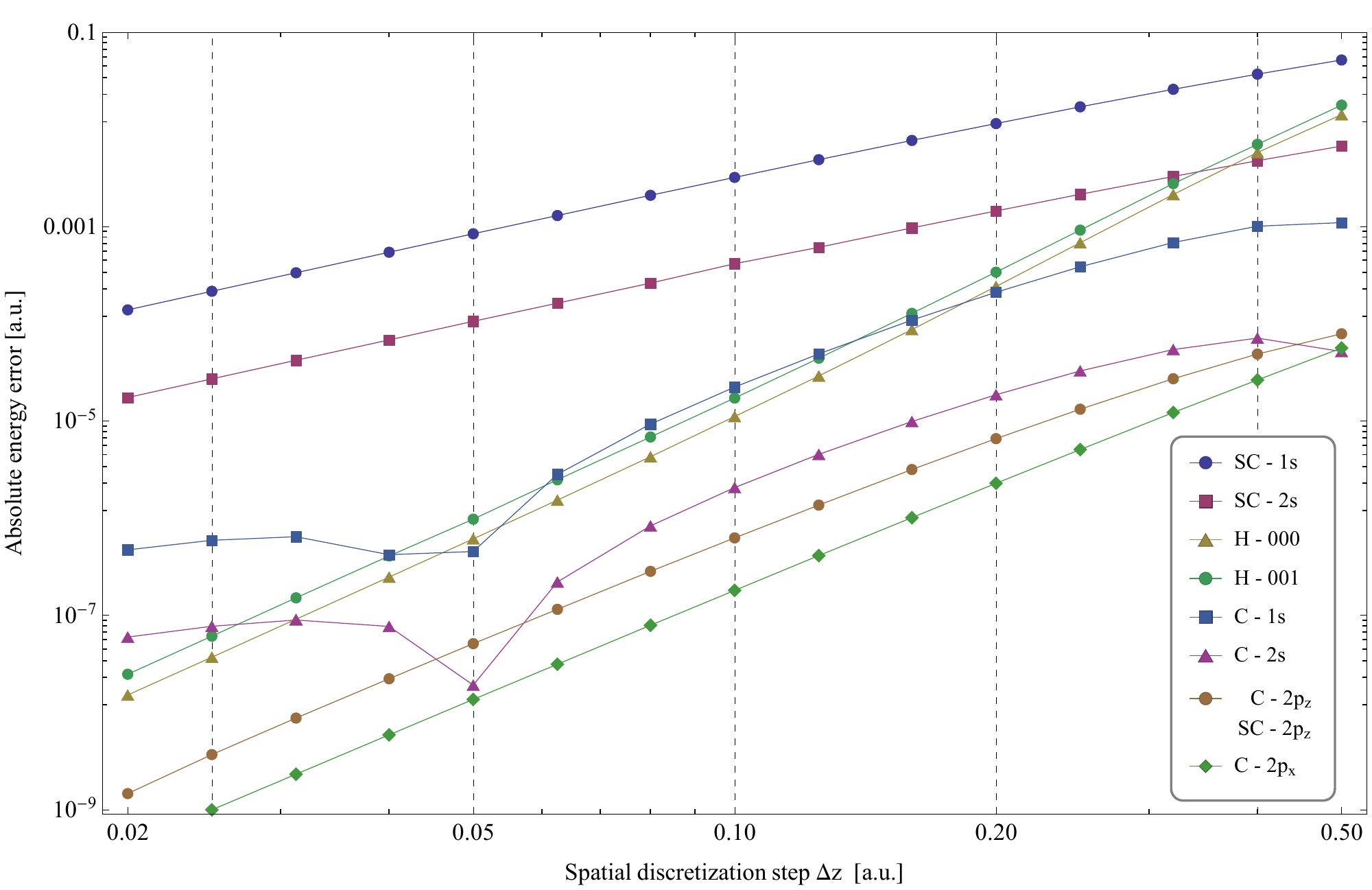} 
\par\end{centering}

\protect\caption{Absolute energy errors of stationary states listed in Table \ref{tab:test_3d_energies_cases},
on a log-log scale. The Coulomb eigenstates were tested with (C) and
without (SC) applying the condition (\ref{eq:cn_3dc_bc_nm_cb}) at
the origin. (In the case of the $2p_{z}$ the error curves are within
line thickness.) Curves for $1s$ and $2s$ show that our hybrid splitting
algorithm decreases the errors with high order even with the singular
Coulomb potential, in the range $0.05\lesssim\Delta z\lesssim0.2$.
\label{fig:test_3d_energies_figure}}
\end{figure}

We computed the eigenenergy values using imaginary time propagation
with the second order hybrid splitting scheme outlined in Section
\ref{sub:splitting_hybrid_scheme}, where the parameter $\Delta t$
was chosen suitably small to minimize the spatial errors related to
the operator splitting. These energy values were also verified with
real time propagation, by comparing the time dependence of the phase
factor of the eigenstates with the exact time dependence of $\Psi_{n}(z,\rho,t)=\psi_{n}(z,\rho)e^{-iE_{n}t}$:
the first two significant digits of the energy error were equal. It
is worth noting that, although we used our hybrid splitting for calculations,
these energy errors are related only to the spatial finite differences
(that is, the 2D Crank-Nicolson scheme). For these results, it was
sufficient to set the ``Crank-Nicolson width'' $L$ just above $\sim1/\Delta z$,
further increase of $L$ did not improve the results.

\begin{table}
\begin{centering}
\begin{tabular*}{1\textwidth}{@{\extracolsep{\fill}}cccccccc}
\hline 
\multicolumn{8}{c}{Absolute energy errors of selected stationary states}\tabularnewline
\hline 
$\Delta z$ & 0.4 & 0.2 & 0.1 & 0.05 & 0.025 & Order & $\text{Energy}_{0}$\tabularnewline
\hline 
\hline 
SC-1s & $3.71\times10^{-2}$ & $1.15\times10^{-2}$ & $3.21\times10^{-3}$ & $8.44\times10^{-4}$ & $2.16\times10^{-4}$ & $1.86$ & $-0.500$\tabularnewline
SC-2s & $4.77\times10^{-3}$ & $1.45\times10^{-3}$ & $4.15\times10^{-4}$ & $1.06\times10^{-4}$ & $2.71\times10^{-5}$ & $1.86$ & $-0.125$\tabularnewline
~~SC-${\rm 2p}{}_{z}$ & $4.88\times10^{-5}$ & $6.57\times10^{-6}$ & $6.25\times10^{-7}$ & $5.10\times10^{-8}$ & $3.68\times10^{-9}$ & $3.42$ & $-0.125$\tabularnewline
C-1s & $1.01\times10^{-3}$ & $2.11\times10^{-4}$ & $2.23\times10^{-5}$ & $4.53\times10^{-7}$ & $5.91\times10^{-7}$ & $2.68$ & $-0.500$\tabularnewline
C-2s & $7.12\times10^{-5}$ & $1.86\times10^{-5}$ & $2.06\times10^{-6}$ & $1.91\times10^{-8}$ & $7.70\times10^{-8}$ & $2.46$ & $-0.125$\tabularnewline
~~C-${\rm 2p}{}_{z}$ & $4.88\times10^{-5}$ & $6.57\times10^{-6}$ & $6.25\times10^{-7}$ & $5.10\times10^{-8}$ & $3.68\times10^{-9}$ & $3.42$ & $-0.125$\tabularnewline
~~C-${\rm 2p}{}_{x}$ & $2.64\times10^{-5}$ & $2.28\times10^{-6}$ & $1.80\times10^{-7}$ & $1.36\times10^{-8}$ & ~$9.93\times10^{-10}$ & $3.67$ & $-0.125$\tabularnewline
H-000 & $5.79\times10^{-3}$ & $2.42\times10^{-4}$ & $1.11\times10^{-5}$ & $6.11\times10^{-7}$ & $3.70\times10^{-8}$ & $4.31$ & $+1.500$\tabularnewline
H-001 & $7.03\times10^{-3}$ & $3.40\times10^{-4}$ & $1.72\times10^{-5}$ & $9.75\times10^{-7}$ & $6.09\times10^{-8}$ & $4.20$ & $+2.500$\tabularnewline
\hline 
\end{tabular*}
\par\end{centering}

\protect\caption{A detailed table of the energy errors at specific values of $\Delta z$
denoted by vertical dashed lines in Figure \ref{fig:test_3d_energies_figure}.
The Order column gives the exponent of $\Delta z$, i.e. the order
of error decrease, valid from step size $\Delta z=0.4$ to $\Delta z=0.025$,
which characterizes the effective accuracy of the method in the given
case. The Coulomb $2p_{z}$ state also has the same error data set
to the first 3 significant digits regardless whether or not the Robin
boundary condition was used. \label{tab:test_3d_energies_table}}
\end{table}

From Figure \ref{fig:test_3d_energies_figure} and Table \ref{tab:test_3d_energies_table}
we can draw several conclusions. In the case of the 3D harmonic oscillator
(a reasonably smooth potential), the method is fourth order accurate
in $\Delta z$, as expected (the effective order is higher than 4
due to the fifth order accurate second derivatives). For the Coulomb
potential, things get considerably more elaborate. In the Soft-Coulomb
case (SC), the energies of the $l=0$ states converge with second
order (more precisely, with 1.86), meaning the method will not be
sufficiently accurate or fast. However, for the true singular Coulomb
case (C), the accuracy of the $l=0$ states drastically improve by
two orders of magnitude around $\Delta z=0.1$. On the other hand,
the complicated step size dependence of the energy error, shown by
the corresponding lines in Figure \ref{fig:test_3d_energies_figure},
does not allow for a true effective order valid in the inspected range.
However, our algorithm does decrease the error with high order (close
to 4) when $0.05\lesssim\Delta z\lesssim0.2$, which is anyhow that
range where the computation runtime is reasonable. The unusual step
size dependence of the energy error below $\Delta z\lesssim0.05$
for data sets C-1s, C-2s is due to finite differencing with high spatial
accuracy applied to a non-analytic problem ($\Psi$ is not continuously
differentiable in the origin), along with the artificially high order
boundary condition.

The test cases with $l>0$ work reasonably well for the $m=0$ configuration
both with C and SC: the SC-${\rm 2p}_{z}$ or the C-${\rm 2p}_{z}$
case has better (relative) accuracy than H-000 but its order is only
$3.42$ as shown in Figure \ref{fig:test_3d_energies_figure}. A slightly
higher order of $3.67$ is achieved for C-${\rm 2p}_{x}$ (the only
test case with $m\neq0$) which has the best accuracy of all computed
eigenenergies in this Section.

In conclusion, the numerical error of the hybrid splitting algorithm
displays a high order scaling with the spatial step size for the stationary
states of the 3D Hydrogen problem.

\subsection{Forced harmonic oscillator\label{sub:test_3d_harmonic}}

In the previous section we focused on the spatial errors in conjunction
with the singular Coulomb potential, now we turn our attention to
the total error of the hybrid splitting algorithm using a smooth time-dependent
potential. This total error is composed typically of several terms
related to the finite differences, to the Padé-approximation, to the
factorization of the exponential operator, and to the short time splitting
of the evolution operator.

\begin{figure}[h]
\begin{centering}
\includegraphics[width=5.5cm]{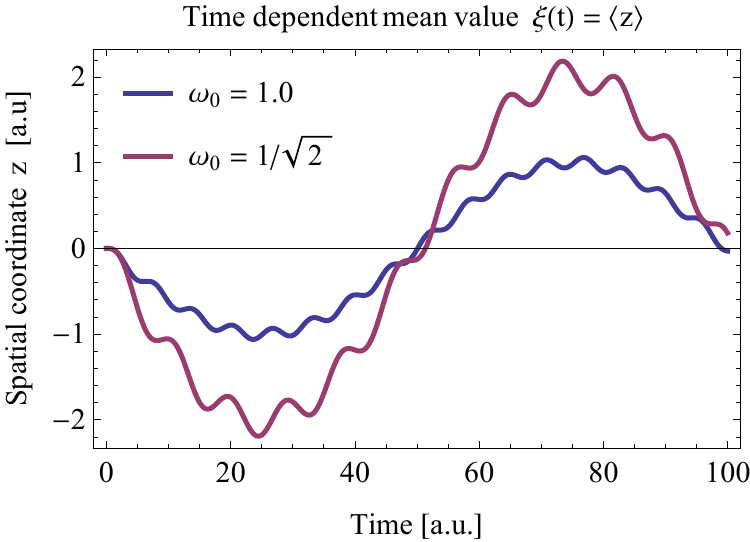}~~~\includegraphics[width=5.5cm]{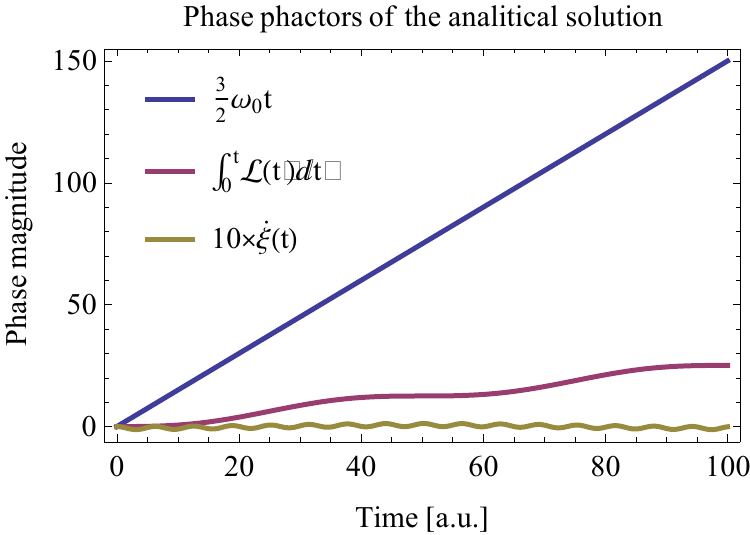} 
\par\end{centering}

\protect\caption{Time evolution of the analytical solution of the forced harmonic oscillator
(\ref{sub:forced_harmonic_oscillator}): the expectation value of
$z$ versus time (left), time dependence of the phase contributions
(right). Parameters are the same as in Table \ref{tab:test_3d_harmonic_table},
if not given explicitly.\label{fig:test_3d_harmonic_support}}
\end{figure}

We use the so called forced harmonic oscillator (FHO) problem as the
test case, defined by the potential $V(z,\rho,t)=\frac{1}{2}\mu\omega_{0}^{2}(z^{2}+\rho^{2})+z{\rm F}\sin\omega_{{\rm F}}t$,
having a known $\Psi^{{\rm A}}(z,\rho,t)$ analytical solution, which
we summarize in \ref{sub:forced_harmonic_oscillator}. Here we only
illustrate this time-dependent analytical solution in Figure \ref{fig:test_3d_harmonic_support}.

For error calculations, we compare the analytical and time-dependent
wavefunctions at a fixed time instant by calculating the so called
mean-square error, which provides information about both the total
phase and the amplitude related errors. We define the mean-square
error by 
\begin{equation}
{\rm err}_{L^{2}}^{2}=2\pi\int\int\rho\left|\Psi^{{\rm A}}(z,\rho)-\Psi(z,\rho)\right|^{2}{\rm d}\rho{\rm d}z\thickapprox2\pi\sum_{i,j}\rho_{j}\left|\Psi_{i,j}^{{\rm A}}-\Psi_{i,j}\right|^{2}.
\end{equation}
We choose the mean-square error as the reference error type because
(according to our experience) it is the most reliable evolution error
quantifier at a fixed time instant $t$.

\begin{table}[h]
\begin{centering}
\begin{tabular}{|c|c|c|}
\hline 
ID & Propagation Method & Evaluations of $S_{2}$\tabularnewline
\hline 
\hline 
S2 & Second order operator hybrid splitting scheme (Sec. \ref{sub:splitting_hybrid_scheme}). & 1\tabularnewline
\hline 
S4E3 & S2 based 4th order iterative splitting scheme using eq. \eqref{eq:splitting_iter_4_u}
with $n=4$. & 3\tabularnewline
\hline 
S6E5 & S2 based 6th order iterative splitting scheme using eq. \eqref{eq:splitting_iter_6_u}
with $n=6$. & 5\tabularnewline
\hline 
S6E9 & S2 based 6th order iterative splitting scheme using eq. \eqref{eq:splitting_iter_4_u}
with $n=6$. & 9\tabularnewline
\hline 
\end{tabular}
\par\end{centering}

\protect\caption{Test case definitions for comparison of the different high order split
operator schemes.\label{tab:test_3d_harmonic_cases}}
\end{table}

We tested our algorithm with different split-operators (listed in
Table \ref{tab:test_3d_harmonic_cases}), and several $\Delta z$
and $\Delta t$ configurations to acquire a top-down view of the error
properties of the hybrid splitting method.

\begin{figure}[h]
\begin{centering}
\includegraphics[width=5.5cm]{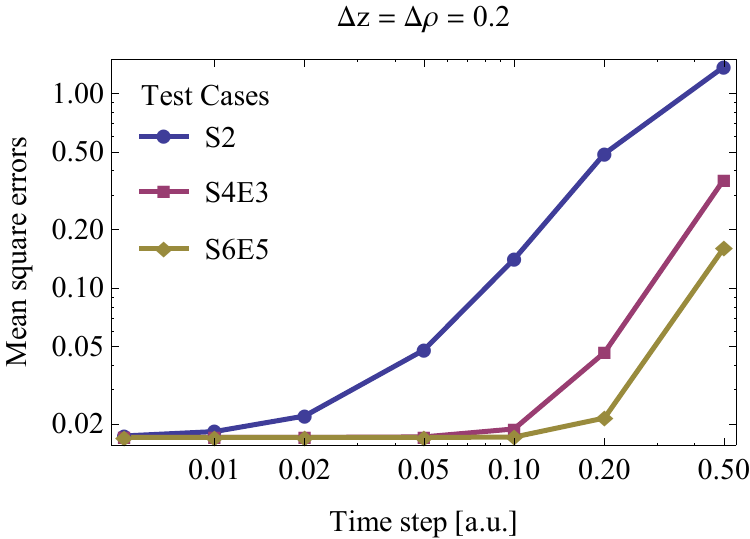}~\includegraphics[width=5.5cm]{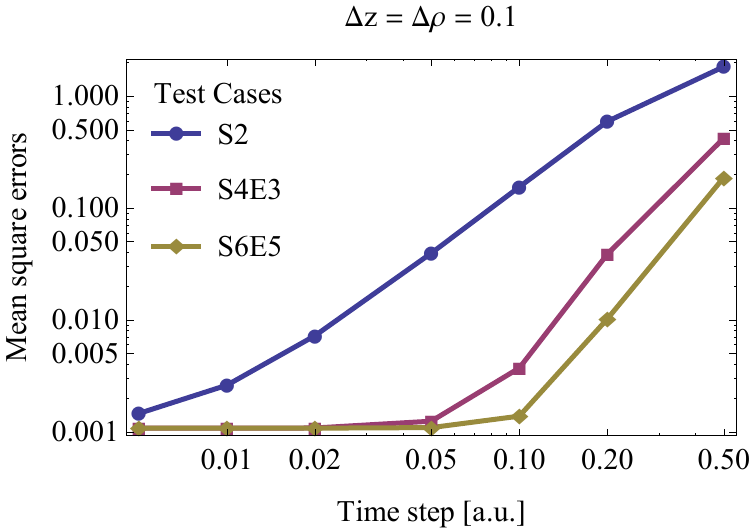}~\includegraphics[width=5.5cm]{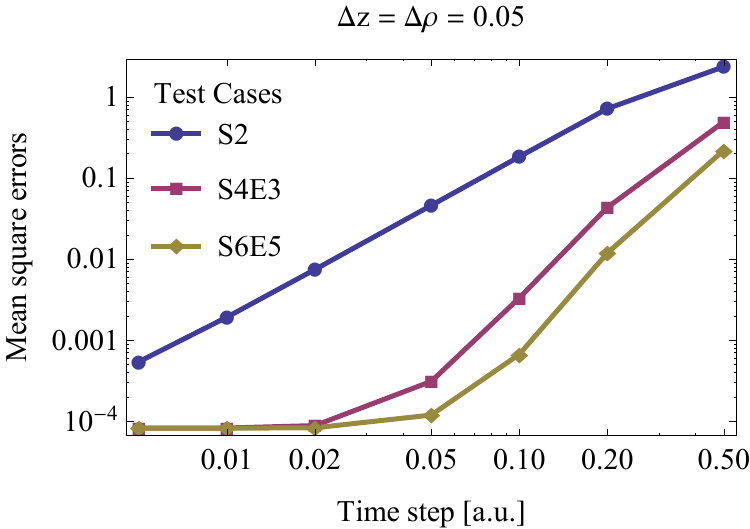} 
\par\end{centering}

\protect\caption{Mean-square errors as the function of the time step, on a log-log
scale, computed with different split-operator methods and spatial
discretization steps, corresponding to Table \ref{tab:test_3d_harmonic_table}.
These plots clearly show the existence of a threshold value of $\Delta t$,
below which the total error is not reduced anymore. \label{fig:test_3d_harmonic_figure}}
\end{figure}

The mean-square error of the different high order split-operators
based on our Crank-Nicolson scheme (CN5) can be found in Table \ref{tab:test_3d_harmonic_table},
and also in Figure \ref{fig:test_3d_harmonic_figure}: the standard
convergence of the S2 scheme is very slow, and there is a drastic
improvement using S4E3 on top of S2. Using either S6E9 or S6E5 does
not yield much gain in accuracy compared to the their higher numerical
costs. Based on these, we propose to use the fourth order split-operator
method S4E3 in accordance with (\ref{eq:splitting_iter_4_u}), along
with our hybrid splitting scheme.

\begin{table}[h]
\begin{centering}
\begin{tabular*}{1\textwidth}{@{\extracolsep{\fill}}cccccccc}
\hline 
\multicolumn{8}{c}{Calculated $L^{2}$ error values at $\Delta z=0.2$}\tabularnewline
\hline 
$\Delta t$ & 0.5 & 0.2 & 0.1 & 0.05 & 0.02 & 0.01 & 0.005\tabularnewline
\hline 
\hline 
S2 & $1.37$ & $4.89\times10^{-1}$ & $1.41\times10^{-1}$ & $4.80\times10^{-2}$ & $2.20\times10^{-2}$ & $1.83\times10^{-2}$ & $1.74\times10^{-2}$\tabularnewline
S4E3 & $3.60\times10^{-1}$ & $4.65\times10^{-2}$ & $1.89\times10^{-2}$ & $1.72\times10^{-2}$ & $1.71\times10^{-2}$ & $1.70\times10^{-2}$ & $1.70\times10^{-2}$\tabularnewline
S6E5 & $1.60\times10^{-1}$ & $2.15\times10^{-2}$ & $1.71\times10^{-2}$ & $1.70\times10^{-2}$ & $1.70\times10^{-2}$ & $1.70\times10^{-2}$ & $1.70\times10^{-2}$\tabularnewline
S6E9 & $1.51\times10^{-1}$ & $2.10\times10^{-2}$ & $1.71\times10^{-2}$ & $1.70\times10^{-2}$ & $1.70\times10^{-2}$ & $1.70\times10^{-2}$ & $1.70\times10^{-2}$\tabularnewline
\hline 
\end{tabular*}
\par\end{centering}

\begin{centering}
\begin{tabular*}{1\textwidth}{@{\extracolsep{\fill}}cccccccc}
\hline 
\multicolumn{8}{c}{Calculated $L^{2}$ error values at $\Delta z=0.1$}\tabularnewline
\hline 
$\Delta t$ & 0.5 & 0.2 & 0.1 & 0.05 & 0.02 & 0.01 & 0.005\tabularnewline
\hline 
\hline 
S2 & 1.83 & $5.95\times10^{-1}$ & $1.54\times10^{-1}$ & $3.95\times10^{-2}$ & $7.22\times10^{-3}$ & $2.61\times10^{-3}$ & $1.46\times10^{-3}$\tabularnewline
S4E3 & $4.20\times10^{-1}$ & $3.87\times10^{-2}$ & $3.72\times10^{-3}$ & $1.24\times10^{-3}$ & $1.08\times10^{-3}$ & $1.08\times10^{-3}$ & $1.08\times10^{-3}$\tabularnewline
S6E5 & $1.87\times10^{-1}$ & $1.02\times10^{-2}$ & $1.38\times10^{-3}$ & $1.09\times10^{-3}$ & $1.08\times10^{-3}$ & $1.08\times10^{-3}$ & $1.08\times10^{-3}$\tabularnewline
S6E9 & $1.76\times10^{-1}$ & $9.50\times10^{-2}$ & $1.36\times10^{-3}$ & $1.09\times10^{-3}$ & $1.08\times10^{-3}$ & $1.08\times10^{-3}$ & $1.08\times10^{-3}$\tabularnewline
\hline 
\end{tabular*}
\par\end{centering}

\begin{centering}
\begin{tabular*}{1\textwidth}{@{\extracolsep{\fill}}cccccccc}
\hline 
\multicolumn{8}{c}{Calculated $L^{2}$ error values at $\Delta z=0.05$}\tabularnewline
\hline 
$\Delta t$ & 0.5 & 0.2 & 0.1 & 0.05 & 0.02 & 0.01 & 0.005\tabularnewline
\hline 
\hline 
S2 & $2.38$ & $7.21\times10^{-1}$ & $1.84\times10^{-1}$ & $4.62\times10^{-2}$ & $7.46\times10^{-3}$ & $1.93\times10^{-3}$ & $5.43\times10^{-4}$\tabularnewline
S4E3 & $4.95\times10^{-1}$ & $4.35\times10^{-2}$ & $3.33\times10^{-3}$ & $3.11\times10^{-4}$ & $8.85\times10^{-5}$ & $8.26\times10^{-5}$ & $8.22\times10^{-5}$\tabularnewline
S6E5 & $2.21\times10^{-1}$ & $1.19\times10^{-2}$ & $6.71\times10^{-4}$ & $1.19\times10^{-4}$ & $8.30\times10^{-5}$ & $8.22\times10^{-5}$ & $8.22\times10^{-5}$\tabularnewline
S6E9 & $2.07\times10^{-1}$ & $1.11\times10^{-2}$ & $6.37\times10^{-4}$ & $1.18\times10^{-4}$ & $8.29\times10^{-5}$ & $8.22\times10^{-5}$ & $8.22\times10^{-5}$\tabularnewline
\hline 
\end{tabular*}
\par\end{centering}

\protect\caption{Mean-square errors of the forced harmonic oscillator with different
$\Delta z$ $(=\Delta\rho)$, $\Delta t$ and split-operator configurations.
All calculations were carried out in a box of $-10\leq z\leq10$ and
$0\leq\rho\leq8$ with propagation parameters $\omega_{{\rm F}}=2\pi/100$,
$F=1$, $\mu=1$, $\omega_{0}=1$, launched from the corresponding
ground state (H-000). All error values are calculated at the time
$t=100$. \label{tab:test_3d_harmonic_table}}
\end{table}

The lines corresponding to the different split-operators in Figure
\ref{fig:test_3d_harmonic_figure} should exhibit the expected power
scaling with $\Delta t$, this is only approximately the case and
only above a threshold $\Delta t$. Below this threshold the total
error is not reduced by decreasing $\Delta t$, because the evolution
error is dominated by the finite differences: the error magnitudes
can even be predicted as the product of the stationary energy error
(Table \ref{tab:test_3d_energies_table}) and the propagation time
interval.

\begin{table}[h]
\begin{centering}
\begin{tabular*}{1\textwidth}{@{\extracolsep{\fill}}cccccccc}
\hline 
\multicolumn{8}{c}{Calculated $L^{2}$ error values of CN3 based S4E3 method}\tabularnewline
\hline 
$\Delta t$ & 0.5 & 0.2 & 0.1 & 0.05 & 0.02 & 0.01 & 0.005\tabularnewline
\hline 
\hline 
$\Delta z=\,\,0.2$ & $7.17\times10^{-1}$ & $4.33\times10^{-1}$ & $4.09\times10^{-1}$ & $4.07\times10^{-1}$ & $4.07\times10^{-1}$ & $4.07\times10^{-1}$ & $4.07\times10^{-1}$\tabularnewline
$\Delta z=\,\,0.1$ & $5.64\times10^{-1}$ & $1.88\times10^{-1}$ & $1.56\times10^{-1}$ & $1.54\times10^{-1}$ & $1.54\times10^{-1}$ & $1.54\times10^{-1}$ & $1.54\times10^{-1}$\tabularnewline
$\Delta z=0.05$ & $5.47\times10^{-2}$ & $9.72\times10^{-2}$ & $5.82\times10^{-2}$ & $5.54\times10^{-2}$ & $5.52\times10^{-2}$ & $5.52\times10^{-2}$ & $5.52\times10^{-2}$\tabularnewline
\hline 
\end{tabular*}
\par\end{centering}

\protect\caption{Mean-square errors of a commonly used propagation scheme, to be compared
with the data of Table \ref{tab:test_3d_harmonic_table}. All the
parameters are the same as in Table \ref{tab:test_3d_harmonic_table}.
Comparison of the last columns shows one, two and three orders of
magnitude accuracy increase in favor of the five point discretization,
as $\Delta z$ is decreased. \label{tab:test_3d_harmonic_table_cn3}}
\end{table}

One objective of our research was to design a real space finite difference
algorithm that achieves fourth order accuracy in the spatial step
size and, most importantly, that is capable to include both the singular
coordinate $\rho=0$ and the singular Coulomb potential directly.
Therefore, it is interesting to globally benchmark the results against
a second order 3-point Crank-Nicolson finite difference scheme (CN3),
which means we degraded the core algorithm to use three-point finite
differences in S2.

We repeated the main tests with this CN3 method combined with the
split-operator configuration S4E3 (discussed above). The CN3 results
are shown in Table \ref{tab:test_3d_harmonic_table_cn3} and are to
be compared to the results with CN5 in Table \ref{tab:test_3d_harmonic_table}.
Both the CN3 and CN5 schemes display the expected order scaling of
the error with $\Delta z$. Comparing the $L^{2}$ error at $\Delta t=0.005$
(which is already below the threshold $\Delta t$), our fourth order
discretization has one, two and three orders of magnitudes smaller
errors at spatial step sizes $\Delta z=0.2,0.1,0.05$, respectively.
Thus, one should always use (at least) the CN5 scheme unless the exotic
nature of the problem prevents high order discretization.

In conclusion, our hybrid splitting propagation method is suitable
for numerical simulations with time-dependent Hamiltonians, and it
is capable of high order scaling with $\Delta t$, once it is incorporated
into the proper high order evolution operator formulae.

\subsection{Hydrogen atom in an external electric field\label{sub:test_3d_coulomb}}

Finally, we conducted a test similar to real world applications by
simulating the hydrogen atom in an external laser field. Now, we compare
our hybrid splitting based CN5 implementation, including Coulomb potential
and the Coulomb boundary condition (CN5-C) against the commonplace
CN3 discretization with the best Soft-Coulomb potential approximation
allowed by the spatial grid (CN3-SC), which is the same approximation
that we tested in Section \ref{sub:test_3d_energies}. We did not
use the full 2D CN3 method, but we applied the hybrid splitting scheme
using the CN3 equations just like the case of the FHO.

The external field was parametrized as $f(t)=F\sin\omega t$. The
simulated time was $t\in[0,100]$, and it consisted of only one field
cycle, by setting $\omega=2\pi/100$ in atomic units. The field amplitude
was $F=0.08$, with simulation box size $z\in[-250,250]$, $\rho\in[0,100]$
to contain the escaping electron waves. The initial state was the
$1s$ ground state, found by imaginary time propagation method to
remove spurious ionization components. We use the S4E3 high order
split operator formula as indicated in the last section.

\begin{figure}[h]
\begin{centering}
\includegraphics[width=15cm]{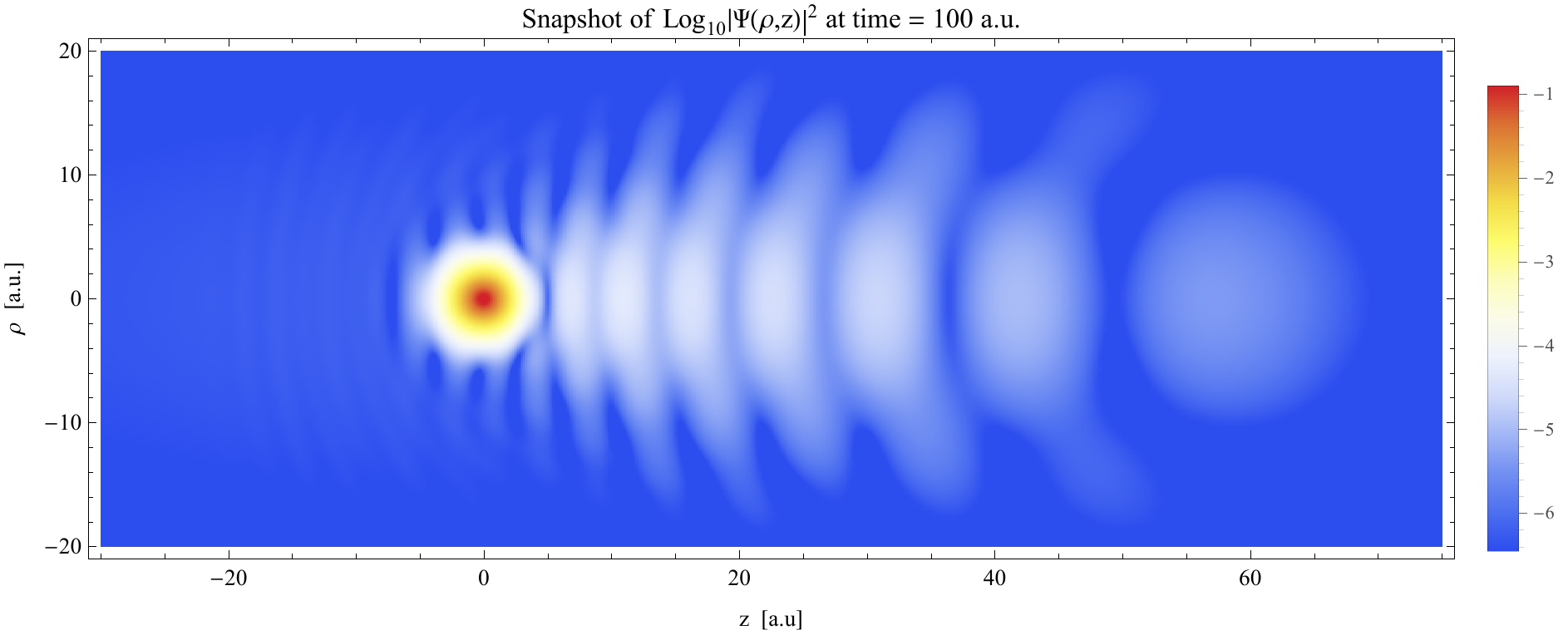} 
\par\end{centering}

\protect\caption{Density plot of the logarithm of the absolute square of the wave function,
in a plane containing the $z$ axis, at the end of the simulation
described in Section \ref{sub:test_3d_coulomb}. Note that the white
waves are 4-5 orders of magnitude smaller than the spherical peak.
\label{fig:test_3d_coulomb_density}}
\end{figure}

We show the result of this simulation in Figure \ref{fig:test_3d_coulomb_density}
by a density plot of the logarithm of the absolute square of the wave
function, in a plane containing the $z$ axis. The white waves on
the right of the spherical peak, bound by the Coulomb potential of
the nucleus, are ca. 4-5 order smaller in density. These waves have
to be computed very accurately, since they contribute the most to
the time dependence of the dipole moment, which in turn is of fundamental
importance regarding the HHG and the creation of attosecond light
pulses.

The tests to compare CN5-C with CN3-SC were run with $\Delta\rho=\Delta z$
and with time step $\Delta t=0.01$ if not indicated explicitly. We
quantified the accuracy of the solutions with the error of the expectation
value $\left\langle z\right\rangle $, i.e. the magnitude of the time-dependent
dipole moment. Unlike the FHO example, analytic solution is not available
this time, so we use a converged solution to determine numerical errors,
that is we compare the results with a much more accurate numerical
solution obtained using smaller $\Delta t$, $\Delta z$ discretization
steps.

\begin{figure}[h]
\begin{centering}
\includegraphics[width=9cm]{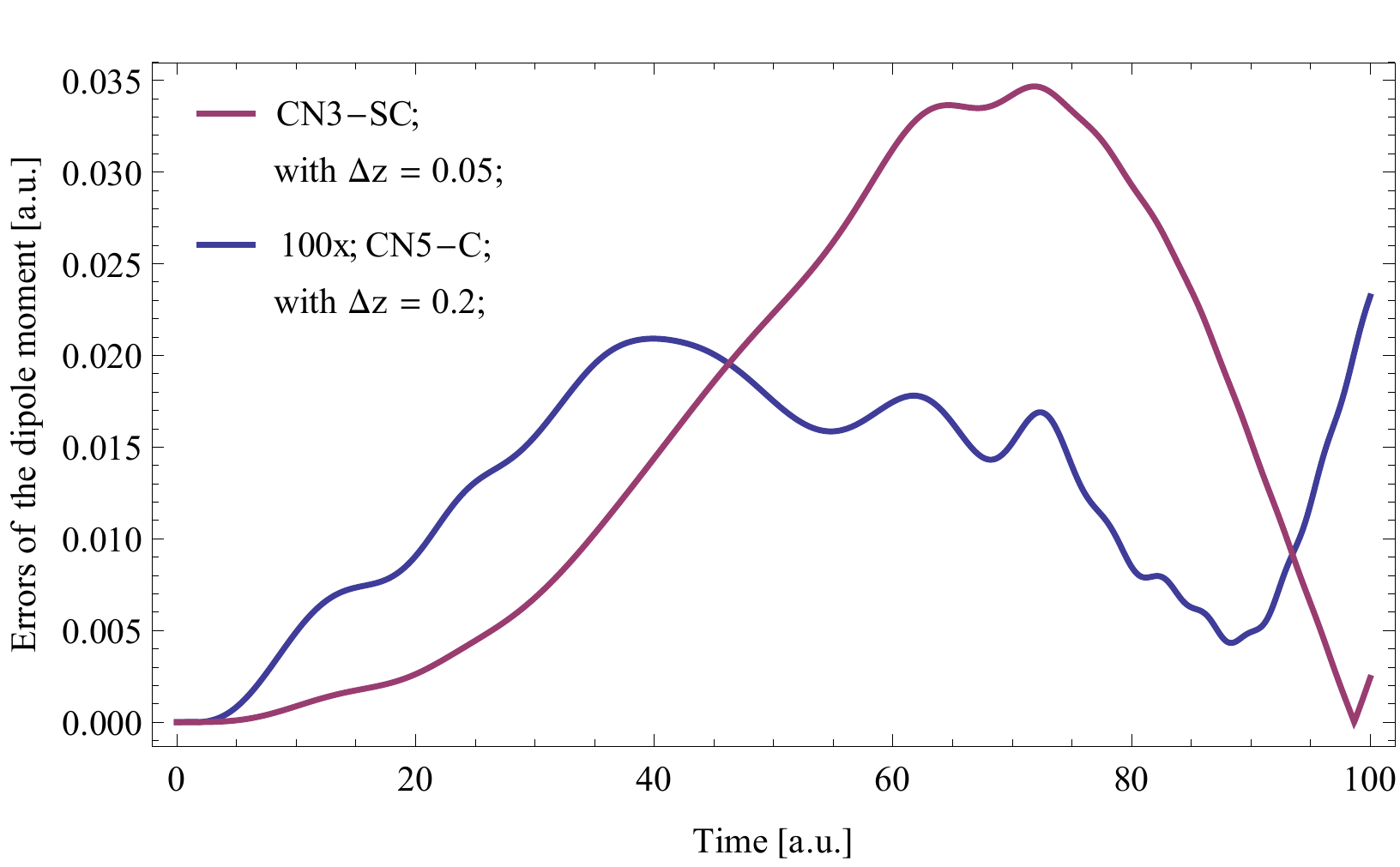} 
\par\end{centering}

\protect\caption{Comparison of the errors of $\left\langle z\right\rangle $, during
the time propagation described in Sec. \ref{sub:test_3d_coulomb},
using the method CN5-C with $\Delta z=0.2$ a.u. (blue line) and using
the method CN3-SC with $\Delta z=0.05$ a.u. (purple line). Despite
the larger spatial step used with CN5-C, its error is two orders of
magnitude smaller, therefore it is shown here with a 100 times magnification.
\label{fig:test_3d_coulomb_meanz}}
\end{figure}

The results, shown in Figure \ref{fig:test_3d_coulomb_meanz} are
as expected: the error of the CN5-C scheme with $\Delta z=0.2$ is
smaller by two orders of magnitude than that of the CN3-SC scheme
with $\Delta z=0.05$. Based on this, we estimate the performance
difference as follows. Due to the second order convergence of the
CN3-SC scheme, $\Delta z\approx0.005$ is needed to achieve the accuracy
of the CN5-C with $\Delta z=0.2$, which means a factor of $1600$
in the number of spatial gridpoints, implying a factor of $800$ -
$1600$ in runtime. That is, the CN5-C scheme is ca. 1000 times more
effective than the CN3-SC. Regarding absolute accuracy, the magnitude
of the error of $\left\langle z\right\rangle $ using CN5-C with $\Delta z=0.2$
is smaller than the $L^{2}$ error of the FHO at the time instant
$t=100$, as can be seen in Table \ref{tab:test_3d_harmonic_table}.

\begin{figure}[h]
\begin{centering}
\includegraphics[width=1\textwidth]{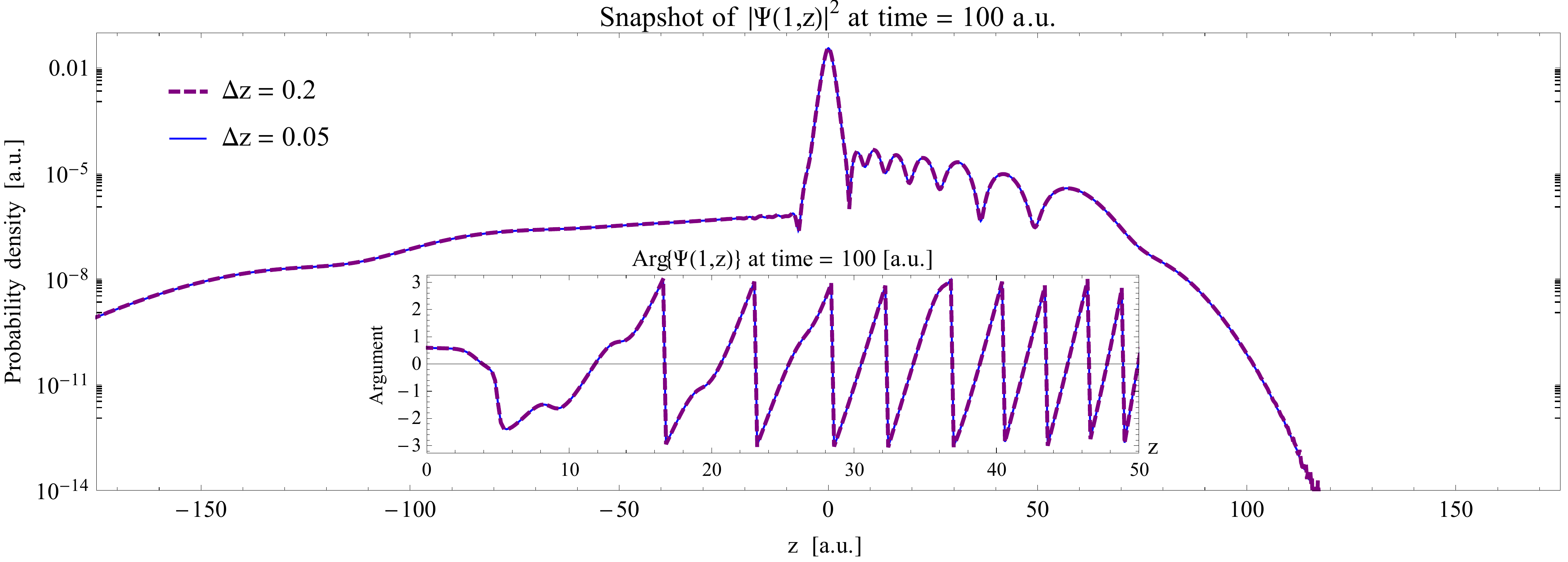} 
\par\end{centering}

\protect\caption{Probability density and phase (in the inset) along a z-line section
at $\rho=1$, at the end of the time propagation described in Sec.
\ref{sub:test_3d_coulomb}, computed using the CN5-C method with two
different values of $\Delta z$, as indicated in the figure. The excellent
fit of the two curves shows that the solution can be considered converged
already at $\Delta z=0.2$. \label{fig:test_3d_coulomb_logz}}
\end{figure}

Regarding the pointwise convergence of the solution with the CN5-C
method, we compare the density and the phase along a $z$-line section
at $\rho=1$, computed with two different values of $\Delta z$. These
plots, shown in Figure \ref{fig:test_3d_coulomb_logz}, convincingly
show that a converged numerical solution is obtained already at $\Delta z=0.2$.
Note that the accuracy of the phase is crucial in strong field physics,
regarding e.g. exit momentum calculations \cite{pfeiffer2012ionizationmomentum,sun2014ionizationmomentum}
in optical tunneling or regarding phase space methods \cite{czirjak2000ionizationwigner,guo2012timeenergyionization,graefe2012quantumphasespace,czirjak2013rescatterentanglement,zagoya2014quantumphasespace,baumann2015strongfieldwigner}
(where usually the Wigner function is computed from the wave function
and this is then further analysed for the physical interpretation
of the process). 

\section{Summary\label{sec:summary_main}}

In this article we presented an algorithm capable of the direct numerical
solution of the three dimensional time dependent Schrödinger equation,
assuming axial symmetry in cylindrical coordinates. The main feature
of the algorithm is that it is capable to accurately handle singular
Coulomb potentials besides any smooth potential of the form $V(z,\rho)$.
The axial symmetry enables a two dimensional grid and the availability
of the Cartesian $z$-axis makes it easy to investigate reduced electron
dynamics.

We choose a high order finite difference representation in the spatial
coordinate domain. We implemented all singularities that can be reduced
to the form $1/\rho$ or $1/r$ via Neumann and Robin boundary conditions
at the $\rho=0$ axis. The accuracy of the algorithm is fourth order
in $\Delta z$, $\Delta\rho$ for smooth potentials and it is close
to fourth order for Coulomb potentials in a restricted discretization
parameter range. We completed this algorithm by a high order scalar
product formula designed for this finite difference representation.

We based our algorithm on the split-operator approximation of the
evolution operator. Due to the non-separability of the Coulomb-problem
in cylindrical coordinates, we constructed a special second order
operator splitting scheme called hybrid splitting method. This splits
the Hamiltonian matrix direction-wise like the traditional methods,
but the innermost region near $\rho=0$ is excluded: here the full
2D Crank-Nicolson equations are used. This means that there are many
decoupled 1D Schrödinger equations in the $z$ direction, and a 2D
Schrödinger equation with a special block pentadiagonal pattern of
the coefficient matrix, which has to be taken advantage of for maximum
efficiency, like in our hybrid splitting solver algorithm. Thread
based parallelization is supported throughout, and we also gave a
way to evaluate the decoupled 1D Schrödinger equations with high spatial
accuracy and efficiency.

We thoroughly investigated the spatial discretization related errors
in an optimal discretization parameter range, determining detailed
accuracy characteristics with or without Coulomb potential. We also
verified the considerably increased accuracy for numerical simulations
forced oscillator. Testing the performance of the 4th and the 6th
order (iterative) split-operator factorizations built from second
order hybrid splitting method, we observed a threshold $\Delta t$
value below which there is no accuracy gain. We concluded that high
order (meaning at least 4th order) split-operator formula should be
used in practice, accompanying the hybrid splitting algorithm.

In order to demonstrate the accuracy and performance of our hybrid
splitting algorithm also in a simulation close to the planned applications,
we computed the solution for a hydrogen atom in a strong time-dependent
electric field of one sine period. The important waves in the probability
density, having an amplitude of $10^{-4}$- $10^{-5}$ relative to
the peak value, were obtained accurately and efficiently.

The hybrid splitting scheme, with some minor modifications of the
algorithm, is also capable to handle single or multiple coupled nonlinear
time dependent Schrödinger equations, such as those arising e.g. in
time-dependent density functional theory \cite{ullrich2014tddftbrief},
time-dependent Hartree-Fock methods \cite{kulander1987tdseionizationhf,pindzola1991tdhfvalidity}
or several other areas of physics.

\section*{Acknowledgements \label{sec:Ack}}

The authors thank W.\ Becker, M.\ G.\ Benedict, P.\ Földi, K.\ Varjú
and S.\ Varró for stimulating discussions. This research has been
granted by the Hungarian Scientific Research Fund OTKA under contract
No. T81364. Partial support by the ELI-ALPS project is also acknowledged.
The ELI-ALPS project (GOP-1.1.1-12/B-2012-000, GINOP-2.3.6-15-2015-00001)
is supported by the European Union and co-financed by the European
Regional Development Fund.

\section*{}

\bibliographystyle{elsarticle-num}
\phantomsection\addcontentsline{toc}{section}{\refname}\bibliography{0Bibliography}

\begin{thebibliography}{10}
\expandafter\ifx\csname url\endcsname\relax
  \def\url#1{\texttt{#1}}\fi
\expandafter\ifx\csname urlprefix\endcsname\relax\def\urlprefix{URL }\fi
\expandafter\ifx\csname href\endcsname\relax
  \def\href#1#2{#2} \def\path#1{#1}\fi

\bibitem{hentschel2001attosecond}
M.~Hentschel, R.~Kienberger, C.~Spielmann, G.~A. Reider, N.~Milosevic,
  T.~Brabec, P.~Corkum, U.~Heinzmann, M.~Drescher, F.~Krausz, Attosecond
  metrology, Nature 414~(6863) (2001) 509--513.

\bibitem{Paul_Science_2001_Atto_pulsetrain}
P.~.~M. Paul, E.~Toma, P.~Breger, G.~Mullot, F.~Aug{\'e}, P.~Balcou, H.~Muller,
  P.~Agostini, Observation of a train of attosecond pulses from high harmonic
  generation, Science 292~(5522) (2001) 1689--1692.

\bibitem{kienberger2002attosecond}
R.~Kienberger, M.~Hentschel, M.~Uiberacker, C.~Spielmann, M.~Kitzler,
  A.~Scrinzi, M.~Wieland, T.~Westerwalbesloh, U.~Kleineberg, U.~Heinzmann,
  et~al., Steering attosecond electron wave packets with light, Science
  297~(5584) (2002) 1144--1148.

\bibitem{drescher2002attosecond}
M.~Drescher, M.~Hentschel, R.~Kienberger, M.~Uiberacker, V.~Yakovlev,
  A.~Scrinzi, T.~Westerwalbesloh, U.~Kleineberg, U.~Heinzmann, F.~Krausz,
  Time-resolved atomic inner-shell spectroscopy, Nature 419~(6909) (2002)
  803--807.

\bibitem{baltuska2003attosecond}
A.~Baltu{\v{s}}ka, T.~Udem, M.~Uiberacker, M.~Hentschel, E.~Goulielmakis,
  C.~Gohle, R.~Holzwarth, V.~Yakovlev, A.~Scrinzi, T.~H{\"a}nsch, et~al.,
  Attosecond control of electronic processes by intense light fields, Nature
  421~(6923) (2003) 611--615.

\bibitem{Krausz_RevModPhys_2009_Attosecond_physics}
F.~Krausz, M.~Ivanov, Attosecond physics, Reviews of Modern Physics 81~(1)
  (2009) 163--234.

\bibitem{McPherson_JOSAB_1987_HHG}
A.~McPherson, G.~Gibson, H.~Jara, U.~Johann, T.~S. Luk, I.~McIntyre, K.~Boyer,
  C.~K. Rhodes, Studies of multiphoton production of vacuum-ultraviolet
  radiation in the rare gases, Journal of the Optical Society of America B
  4~(4) (1987) 595--601.

\bibitem{Ferray_JPhysB_1988_HHG}
M.~Ferray, A.~L'Huillier, X.~Li, L.~Lompre, G.~Mainfray, C.~Manus,
  Multiple-harmonic conversion of 1064 nm radiation in rare gases, Journal of
  Physics B: Atomic, Molecular and Optical Physics 21~(3) (1988) L31.

\bibitem{Farkas_PhysLettA_1992_AttoPulse}
G.~Farkas, C.~T{\'o}th, Proposal for attosecond light pulse generation using
  laser induced multiple-harmonic conversion processes in rare gases, Physics
  Letters A 168~(5) (1992) 447--450.

\bibitem{Harris_OptCom_1993_HHG_Atto}
S.~Harris, J.~Macklin, T.~H{\"a}nsch, Atomic scale temporal structure inherent
  to high-order harmonic generation, Optics communications 100~(5-6) (1993)
  487--490.

\bibitem{Uiberacker_Nature_2007}
M.~Uiberacker, T.~Uphues, M.~Schultze, A.~J. Verhoef, V.~Yakovlev, M.~F. Kling,
  J.~Rauschenberger, N.~M. Kabachnik, H.~Schr{\"o}der, M.~Lezius, et~al.,
  Attosecond real-time observation of electron tunnelling in atoms, Nature
  446~(7136) (2007) 627--632.

\bibitem{Haessler_NatPhys_2010}
S.~Haessler, J.~Caillat, W.~Boutu, C.~Giovanetti-Teixeira, T.~Ruchon,
  T.~Auguste, Z.~Diveki, P.~Breger, A.~Maquet, B.~Carr{\'e}, et~al., Attosecond
  imaging of molecular electronic wavepackets, Nature Physics 6~(3) (2010)
  200--206.

\bibitem{pfeiffer2012ionizationmomentum}
A.~N. Pfeiffer, C.~Cirelli, A.~S. Landsman, M.~Smolarski, D.~Dimitrovski, L.~B.
  Madsen, U.~Keller, Probing the longitudinal momentum spread of the electron
  wave packet at the tunnel exit, Physical Review Letters 109~(8) (2012)
  083002.

\bibitem{Shafir_Nature_2012}
D.~Shafir, H.~Soifer, B.~D. Bruner, M.~Dagan, Y.~Mairesse, S.~Patchkovskii,
  M.~Y. Ivanov, O.~Smirnova, N.~Dudovich, Resolving the time when an electron
  exits a tunnelling barrier, Nature 485~(7398) (2012) 343--346.

\bibitem{Schultze_Science_2010}
M.~Schultze, M.~Fie{\ss}, N.~Karpowicz, J.~Gagnon, M.~Korbman, M.~Hofstetter,
  S.~Neppl, A.~L. Cavalieri, Y.~Komninos, T.~Mercouris, et~al., Delay in
  photoemission, Science 328~(5986) (2010) 1658--1662.

\bibitem{Cavalieri_Nature_2007}
A.~L. Cavalieri, N.~M{\"u}ller, T.~Uphues, V.~S. Yakovlev, A.~Baltu{\v{s}}ka,
  B.~Horvath, B.~Schmidt, L.~Bl{\"u}mel, R.~Holzwarth, S.~Hendel, et~al.,
  Attosecond spectroscopy in condensed matter, Nature 449~(7165) (2007)
  1029--1032.

\bibitem{BOOK_QUANTUM_IMAGING_BANDRAUK_2011}
A.~D. Bandrauk, M.~Ivanov, Quantum Dynamic Imaging: Theoretical and Numerical
  Methods, Springer Science \& Business Media, 2011.

\bibitem{Keldysh_JETP_1965}
L.~Keldysh, Ionization in the field of a strong electromagnetic wave, Soviet
  Physics JETP 20~(5) (1965) 1307--1314.

\bibitem{Corkum_PRL_1993}
P.~B. Corkum, Plasma perspective on strong field multiphoton ionization,
  Physical Review Letters 71~(13) (1993) 1994.

\bibitem{eckle2008attosecond}
P.~Eckle, A.~Pfeiffer, C.~Cirelli, A.~Staudte, R.~D{\"o}rner, H.~Muller,
  M.~B{\"u}ttiker, U.~Keller, Attosecond ionization and tunneling delay time
  measurements in helium, Science 322~(5907) (2008) 1525--1529.

\bibitem{Lewenstein_PRA_1994}
M.~Lewenstein, P.~Balcou, M.~Y. Ivanov, A.~L'Huillier, Auillier, P.~B. Corkum,
  Theory of high-harmonic generation by low-frequency laser fields, Physical
  Review A 49~(3) (1994) 2117.

\bibitem{ivanov2005strongfield}
M.~Y. Ivanov, M.~Spanner, O.~Smirnova, Anatomy of strong field ionization,
  Journal of Modern Optics 52~(2-3) (2005) 165--184.

\bibitem{gordon2005tdsecoulomb}
A.~Gordon, R.~Santra, F.~X. K{\"a}rtner, Role of the coulomb singularity in
  high-order harmonic generation, Physical Review A 72~(6) (2005) 063411.

\bibitem{tong1997tdsehhgspectral}
X.-M. Tong, S.-I. Chu, Theoretical study of multiple high-order harmonic
  generation by intense ultrashort pulsed laser fields: A new generalized
  pseudospectral time-dependent method, Chemical Physics 217~(2) (1997)
  119--130.

\bibitem{bachau2001tdsebspline}
H.~Bachau, E.~Cormier, P.~Decleva, J.~Hansen, F.~Martin, Applications of
  b-splines in atomic and molecular physics, Reports on Progress in Physics
  64~(12) (2001) 1815.

\bibitem{cormier1997tdsebspline}
E.~Cormier, P.~Lambropoulos, Above-threshold ionization spectrum of hydrogen
  using b-spline functions, Journal of Physics B: Atomic, Molecular and Optical
  Physics 30~(1) (1997) 77.

\bibitem{muller1999tdseefficient}
H.~Muller, An efficient propagation scheme for the time-dependent
  {Schr{\"o}dinger} equation in the velocity gauge, Laser Physics 9 (1999)
  138--148.

\bibitem{bauer2006tdseqprop}
D.~Bauer, P.~Koval, Qprop: A {Schr{\"o}dinger}-solver for intense laser--atom
  interaction, Computer Physics Communications 174~(5) (2006) 396--421.

\bibitem{smyth1998tdsehelium}
E.~S. Smyth, J.~S. Parker, K.~T. Taylor, Numerical integration of the
  time-dependent {Schr{\"o}dinger} equation for laser-driven helium, Computer
  Physics Communications 114~(1) (1998) 1--14.

\bibitem{argenti2013photoionization}
L.~Argenti, R.~Pazourek, J.~Feist, S.~Nagele, M.~Liertzer, E.~Persson,
  J.~Burgd{\"o}rfer, E.~Lindroth, Photoionization of helium by attosecond
  pulses: Extraction of spectra from correlated wave functions, Physical Review
  A 87~(5) (2013) 053405.

\bibitem{schneider2006tdsesolver}
B.~I. Schneider, L.~A. Collins, S.~Hu, Parallel solver for the time-dependent
  linear and nonlinear {Schr{\"o}dinger} equation, Physical Review E 73~(3)
  (2006) 036708.

\bibitem{kolakowska1998excitation}
A.~Ko{\l}akowska, M.~Pindzola, F.~Robicheaux, D.~Schultz, J.~Wells, Excitation
  and charge transfer in proton-hydrogen collisions, Physical Review A 58~(4)
  (1998) 2872.

\bibitem{christov1997tdsehhg}
I.~P. Christov, M.~M. Murnane, H.~C. Kapteyn, High-harmonic generation of
  attosecond pulses in the single-cycle regime, Physical Review Letters 78~(7)
  (1997) 1251.

\bibitem{hu2004tdsehhg}
S.~Hu, L.~Collins, Intense laser-induced recombination: The inverse
  above-threshold ionization process, Physical Review A 70~(1) (2004) 013407.

\bibitem{wells1996tdsenumerical}
J.~Wells, D.~Schultz, P.~Gavras, M.~Pindzola, Numerical solution of the
  time-dependent {Schr{\"o}dinger} equation for intermediate-energy collisions
  of antiprotons with hydrogen, Physical Review A 54~(1) (1996) 593.

\bibitem{gordon2006tdsecoulomb}
A.~Gordon, C.~Jirauschek, F.~X. K{\"a}rtner, Numerical solver of the
  time-dependent {Schr{\"o}dinger} equation with coulomb singularities,
  Physical Review A 73~(4) (2006) 042505.

\bibitem{gainullin2015tdsesolver}
I.~Gainullin, M.~Sonkin, High-performance parallel solver for 3d time-dependent
  {Schr{\"o}dinger} equation for large-scale nanosystems, Computer Physics
  Communications 188 (2015) 68--75.

\bibitem{leforestier1991tdsecomparison}
C.~Leforestier, R.~Bisseling, C.~Cerjan, M.~Feit, R.~Friesner, A.~Guldberg,
  A.~Hammerich, G.~Jolicard, W.~Karrlein, H.-D. Meyer, et~al., A comparison of
  different propagation schemes for the time dependent {Schr{\"o}dinger}
  equation, Journal of Computational Physics 94~(1) (1991) 59--80.

\bibitem{iitaka1994tdsemsd4}
T.~Iitaka, Solving the time-dependent {Schr{\"o}dinger} equation numerically,
  Physical Review E 49~(5) (1994) 4684.

\bibitem{castro2004kohnshampropagators}
A.~Castro, M.~A. Marques, A.~Rubio, Propagators for the time-dependent
  {Kohn--Sham} equations, The Journal of chemical physics 121~(8) (2004)
  3425--3433.

\bibitem{blanes2000tdsesplitting}
S.~Blanes, P.~Moan, Splitting methods for the time-dependent {Schr{\"o}dinger}
  equation, Physics Letters A 265~(1) (2000) 35--42.

\bibitem{chin2001splittinggrad}
S.~A. Chin, C.-R. Chen, Fourth order gradient symplectic integrator methods for
  solving the time-dependent {Schr{\"o}dinger} equation, The Journal of
  Chemical Physics 114~(17) (2001) 7338--7341.

\bibitem{BOOK_NUMERICAL_RECIPIES_2007}
W.~H. Press, S.~A. Teukolsky, W.~T. Vetterling, B.~P. Flannery, Numerical
  Recipes 3rd Edition: The Art of Scientific Computing, 3rd Edition, Cambridge
  University Press, 2007.

\bibitem{puzynin1999tdsepade}
I.~Puzynin, A.~Selin, S.~Vinitsky, A high-order accuracy method for numerical
  solving of the time-dependent {Schr{\"o}dinger} equation, Computer Physics
  Communications 123~(1) (1999) 1--6.

\bibitem{wang2005nltdsedifferences}
H.~Wang, Numerical studies on the split-step finite difference method for
  nonlinear {Schr{\"o}dinger} equations, Applied Mathematics and Computation
  170~(1) (2005) 17--35.

\bibitem{shen2013tdsesolving}
J.~Shen, E.~Wei, Z.~Huang, M.~Chen, X.~Wu, High-order symplectic fdtd scheme
  for solving a time-dependent {Schr{\"o}dinger} equation, Computer Physics
  Communications 184~(3) (2013) 480--492.

\bibitem{BOOK_ATOMS_MOLECULES_BRANSDEN_2003}
B.~H. Bransden, C.~J. Joachain, Physics of Atoms and Molecules, Pearson
  Education India, 2003.

\bibitem{BOOK_QUANTUM_COHEN_1997}
C.~Cohen-Tannoudji, B.~Diu, F.~Lalo{\"e}, Quantum Mechanics, Quantum Mechanics,
  Wiley, 1977.

\bibitem{BOOK_QUANTUM_GRIFFITS_2005}
D.~J. Griffiths, Introduction to Quantum Mechanics, Pearson Education India,
  2005.

\bibitem{puzynin2000tdsemagnus}
I.~Puzynin, A.~Selin, S.~Vinitsky, Magnus-factorized method for numerical
  solving the time-dependent {Schr{\"o}dinger} equation, Computer Physics
  Communications 126~(1) (2000) 158--161.

\bibitem{blanes2009magnus}
S.~Blanes, F.~Casas, J.~Oteo, J.~Ros, The magnus expansion and some of its
  applications, Physics Reports 470~(5) (2009) 151--238.

\bibitem{dijk2007tdseaccurate}
W.~van Dijk, F.~Toyama, Accurate numerical solutions of the time-dependent
  {Schr{\"o}dinger} equation, Physical Review E 75~(3) (2007) 36707.

\bibitem{varah1972blocktridiagonal}
J.~M. Varah, On the solution of block-tridiagonal systems arising from certain
  finite-difference equations, Mathematics of Computation 26~(120) (1972)
  859--868.

\bibitem{wilcox1967exponential}
R.~Wilcox, Exponential operators and parameter differentiation in quantum
  physics, Journal of Mathematical Physics 8~(4) (1967) 962--982.

\bibitem{mclachlan2002splitting}
R.~I. McLachlan, G.~R.~W. Quispel, Splitting methods, Acta Numerica 11 (2002)
  341--434.

\bibitem{yazici2010operator}
Y.~Yazici, Operator splitting methods for differential equations, Ph.D. thesis,
  Izmir Institute of Technology (2010).

\bibitem{strang1968splitting}
G.~Strang, On the construction and comparison of difference schemes, SIAM
  Journal on Numerical Analysis 5~(3) (1968) 506--517.

\bibitem{chin1997splittingsymplectic}
S.~A. Chin, Symplectic integrators from composite operator factorizations,
  Physics Letters A 226~(6) (1997) 344--348.

\bibitem{suzuki1995splittinghybrid}
M.~Suzuki, Hybrid exponential product formulas for unbounded operators with
  possible applications to monte carlo simulations, Physics Letters A 201~(5)
  (1995) 425--428.

\bibitem{bandrauk1992splitting}
A.~D. Bandrauk, H.~Shen, Higher order exponential split operator method for
  solving time-dependent {Schr{\"o}dinger} equations, Canadian Journal of
  Chemistry 70~(2) (1992) 555--559.

\bibitem{bandrauk2013splitting}
A.~D. Bandrauk, H.~Lu, Exponential propagators (integrators) for the
  time-dependent {Schr{\"o}dinger} equation, Journal of Theoretical and
  Computational Chemistry 12~(06) (2013) 1340001.

\bibitem{bandrauk1993splitting}
A.~D. Bandrauk, H.~Shen, Exponential split operator methods for solving coupled
  time-dependent {Schr{\"o}dinger} equations, The Journal of chemical physics
  99~(2) (1993) 1185--1193.

\bibitem{chin2009imagsplit}
S.~A. Chin, S.~Janecek, E.~Krotscheck, Any order imaginary time propagation
  method for solving the {Schr{\"o}dinger} equation, Chemical Physics Letters
  470~(4) (2009) 342--346.

\bibitem{lehtovaara2007tdseimaginary}
L.~Lehtovaara, J.~Toivanen, J.~Eloranta, Solution of time-independent
  {Schr{\"o}dinger} equation by the imaginary time propagation method, Journal
  of Computational Physics 221~(1) (2007) 148--157.

\bibitem{muga2004absorbingpotentials}
J.~G. Muga, J.~Palao, B.~Navarro, I.~Egusquiza, Complex absorbing potentials,
  Physics Reports 395~(6) (2004) 357--426.

\bibitem{karawia2006solution5diagonal}
A.~Karawia, A computational algorithm for solving periodic penta-diagonal
  linear systems, Applied Mathematics and Computation 174~(1) (2006) 613--618.

\bibitem{sun2014ionizationmomentum}
X.~Sun, M.~Li, J.~Yu, Y.~Deng, Q.~Gong, Y.~Liu, Calibration of the initial
  longitudinal momentum spread of tunneling ionization, Physical Review A
  89~(4) (2014) 045402.

\bibitem{czirjak2000ionizationwigner}
A.~Czirj{\'a}k, R.~Kopold, W.~Becker, M.~Kleber, W.~Schleich, The {Wigner}
  function for tunneling in a uniform static electric field, Optics
  communications 179~(1) (2000) 29--38.

\bibitem{guo2012timeenergyionization}
L.~Guo, S.~Han, J.~Chen, Time-energy analysis of above-threshold ionization in
  few-cycle laser pulses, Physical Review A 86~(5) (2012) 053409.

\bibitem{graefe2012quantumphasespace}
S.~Gr{\"a}fe, J.~Doose, J.~Burgd{\"o}rfer, Quantum phase-space analysis of
  electronic rescattering dynamics in intense few-cycle laser fields, Journal
  of Physics B: Atomic, Molecular and Optical Physics 45~(5) (2012) 055002.

\bibitem{czirjak2013rescatterentanglement}
A.~Czirj{\'a}k, S.~Majorosi, J.~Kov{\'a}cs, M.~G. Benedict, Emergence of
  oscillations in quantum entanglement during rescattering, Physica Scripta
  2013~(T153) (2013) 014013.

\bibitem{zagoya2014quantumphasespace}
C.~Zagoya, J.~Wu, M.~Ronto, D.~Shalashilin, C.~F. de~Morisson~Faria, Quantum
  and semiclassical phase-space dynamics of a wave packet in strong fields
  using initial-value representations, New Journal of Physics 16~(10) (2014)
  103040.

\bibitem{baumann2015strongfieldwigner}
C.~Baumann, H.-J. Kull, G.~Fraiman, {Wigner} representation of ionization and
  scattering in strong laser fields, Physical Review A 92~(6) (2015) 063420.

\bibitem{ullrich2014tddftbrief}
C.~A. Ullrich, Z.-h. Yang, A brief compendium of time-dependent density
  functional theory, Brazilian Journal of Physics 44~(1) (2014) 154--188.

\bibitem{kulander1987tdseionizationhf}
K.~C. Kulander, Time-dependent {Hartree-Fock} theory of multiphoton ionization:
  Helium, Physical Review A 36~(6) (1987) 2726.

\bibitem{pindzola1991tdhfvalidity}
M.~Pindzola, D.~Griffin, C.~Bottcher, Validity of time-dependent {Hartree-Fock}
  theory for the multiphoton ionization of atoms, Physical Review Letters
  66~(18) (1991) 2305.

\bibitem{moyer2004tdsenumerovtransparent}
C.~A. Moyer, Numerov extension of transparent boundary conditions for the
  {Schr{\"o}dinger} equation in one dimension, American Journal of Physics
  72~(3) (2004) 351--358.

\bibitem{husimi1953miscellanea}
K.~Husimi, Miscellanea in elementary quantum mechanics, ii, Progress of
  Theoretical Physics 9~(4) (1953) 381--402.

\end{thebibliography}

\appendix

\section{Approximating the inner product\label{sub:dot_product_3dc}}

In this Section we propose a solution to the problem of the discrete
inner product formula exposed in Section \ref{sec:crank_nicolson_3dc}.

We seek a discretized representation of the inner product formula
in the cylindrical coordinate system as

\begin{equation}
\left\langle \Phi|\Psi\right\rangle =2\pi\int_{-\infty}^{+\infty}\int_{0}^{\infty}\rho\,\,\Phi^{*}(z,\rho)\Psi(z,\rho)\text{d}\rho\text{d}z=\sum_{i,j}c_{i,j}\Phi_{i,j}^{*}\Psi_{i,j}.\label{eq:dot_3dc_null}
\end{equation}
The naive approach with coefficients $c_{i,j}=2\pi\rho_{j}\Delta\rho\Delta z$
causes inaccuracy which originates from the $\rho=0$ edge and its
neighborhood only, because the formula with this particular $c_{i,j}$
has exponential convergence at the box boundaries \cite{BOOK_NUMERICAL_RECIPIES_2007}.

Besides the accuracy, the conservation of a discretized scheme of
form (\ref{eq:dot_3dc_null}) is also an issue: given a discrete Hamiltonian
matrix $H$ in the Padé-approximation then the time evolution will
be unitary with respect to the inner product $\sum_{i,j}c_{i,j}\Phi_{i,j}^{*}\Psi_{i,j}$
only if $H$ is self-adjoint. Unfortunately, because of the Neumann
and Robin boundary conditions (\ref{eq:cn_3dc_full_nm_cb}) the Hamiltonian
matrix $H$ does not exist on the $\rho=0$ line, therefore there
is no standard norm of form $\sum_{i,j}c_{i,j}\Psi_{i,j}^{*}\Psi_{i,j}$
which is perfectly conserved by the numerical scheme.

Therefore, our aim is to approximate (\ref{eq:dot_3dc_null}) with
an order that is higher than the order of the finite difference scheme.
To achieve this goal, we used Lagrange \cite{BOOK_NUMERICAL_RECIPIES_2007}
interpolating polynomial $p_{i,j}(\rho)$ defined on points $(\rho_{j},Q_{i,j})$,
$(\rho_{j+1},Q_{i,j+1})$, $(\rho_{j+2},Q_{i,j+2})$, $(\rho_{j+3},Q_{i,j+3})$,
$(\rho_{j+4},Q_{i,j+4})$, $(\rho_{j+5},Q_{i,j+5})$, $(\rho_{j+6},Q_{i,j+6})$
for a given $z_{i}$ line, where $Q{}_{i,j}$ is the integrand in
$\left\langle \Phi|\Psi\right\rangle =\iint Q(z,\rho)\text{d}\rho\text{d}z$.
We use an elementary integral formula between $\rho_{j}$ and $\rho_{j+1}$:
\begin{align}
\intop_{\rho_{j}}^{\rho_{j+1}}Q(z,\rho)\text{d}\rho= & \left[\frac{19087}{60480}Q_{i,j}+\frac{2713}{2520}Q_{i,j+1}-\frac{15487}{20160}Q_{i,j+2}+\frac{586}{945}Q_{i,j+3}-\frac{6737}{20160}Q_{i,j+4}\right.\nonumber \\
\, & \left.\,\,\frac{263}{2520}Q_{i,j+5}-\frac{863}{60480}Q_{i,j+6}\right]\cdot\Delta\rho+O\left(\Delta\rho^{8}\right).\label{eq:dot_3dc_formula_6_one}
\end{align}
We sum up this for all $i,j$ points, and utilize the boundary conditions
for $j\geq N_{\rho}$, then we arrive at the following integral formula,
which is our choice to approximate the scalar product (\ref{eq:dot_3dc_null}):

\begin{align}
\left\langle \Phi|\Psi\right\rangle =\sum_{i=0}^{N_{z}} & \left[\frac{19087}{60480}\rho_{0}\Phi_{i,0}^{*}\Psi_{i,0}+\frac{84199}{60480}\rho_{1}\Phi_{i,1}^{*}\Psi_{i,1}+\frac{18869}{30240}\rho_{2}\Phi_{i,2}^{*}\Psi_{i,2}+\frac{37621}{30240}\rho_{3}\Phi_{i,3}^{*}\Psi_{i,3}+\right.\nonumber \\
\, & \left.\,\,\frac{55031}{60480}\rho_{4}\Phi_{i,4}^{*}\Psi_{i,4}+\frac{61343}{60480}\rho_{5}\Phi_{i,5}^{*}\Psi_{i,5}+\sum_{j=6}^{N_{\rho}}\rho_{j}\Phi_{i,j}^{*}\Psi_{i,j}\right]\cdot2\pi\Delta\rho\Delta z.\label{eq:dot_3dc_formula_6_full}
\end{align}
This achieves the high integration accuracy (it is exact for a polynomial
of $\rho$ up to degree 6), which is needed: the computed norm variations
become proportional to $\Delta\rho^{4}$, which is consistent with
the accuracy of the spatial finite differences. The constructed Crank-Nicolson
scheme stayed stable in our simulations.

\section{Numerov z-line propagator algorithm\label{sub:numerov_z_line}}

We have constructed an efficient way of evaluating $e^{-i\Delta t\, H_{z}}$
line-by-line, based on the second order Crank-Nicolson algorithm with
Numerov-extension \cite{moyer2004tdsenumerovtransparent}, which also
provides at least fourth order accuracy in $\Delta z$, and it reduces
the numerical costs because it uses three point finite differences
and tridiagonal equations instead of five point differences and pentadiagonal
equations. We will also outline an optimization technique for 1D Crank-Nicolson
schemes which is makes the computations even more efficient when they
are applied to 2D problems.

Let us fix the value of coordinate $\rho$ (i.e we choose $j=\text{const}$)
and we denote $\Psi_{i}(t_{k})=\Psi_{i,j}(t_{k})$. Here we consider
the Hamiltonian in the form of $H_{z}=\beta\partial_{z}^{2}+V(z,t)$
along this line, since the procedure below allows to include a potential
of this form, which is however not present in our hybrid splitting
scheme. Again, we start with the second order approximation of the
exponential operator $e^{-i\Delta t\, H_{z}}$ as in (\ref{eq:time_pade_cn_2})
\begin{equation}
\left(1+\alpha\beta\partial_{z}^{2}+\alpha V\right)\Psi(t_{k+1})=\left(1-\alpha\beta\partial_{z}^{2}-\alpha V\right)\Psi(t_{k})\label{eq:numerov_z_begin}
\end{equation}
with $\alpha=i\Delta t/2$ , $V=V(z,t_{k+1/2})$. We will use a discretized
Laplacian $L_{z}$ based on the standard three point finite difference
method \cite{BOOK_NUMERICAL_RECIPIES_2007}, however, we will also
need the leading term of the error: 
\begin{equation}
L_{z}\Psi_{i}=\frac{\Psi_{i-1}-2\Psi_{i}+\Psi_{i+1}}{\Delta z^{2}}\approx\left.\frac{\partial^{2}\Psi}{\partial z^{2}}\right|_{z_{i}}-\frac{\Delta z^{2}}{12}\left.\frac{\partial^{4}\Psi}{\partial z^{4}}\right|_{z_{i}}+O(\Delta z^{4}).\label{eq:numerov_z_laplacian_3}
\end{equation}
To proceed, we introduce the auxiliary variable $Y(t_{k})=\Psi(t_{k+1})+\Psi(t_{k})$
and rewrite the equation (\ref{eq:numerov_z_begin}) as 
\begin{equation}
\left(\alpha\beta\partial_{z}^{2}\right)Y(t_{k})=2\Psi(t_{k})-\left(1+\alpha V\right)Y(t_{k}).\label{eq:numerov_z_contiuous}
\end{equation}
Then, we discretize the equation (\ref{eq:numerov_z_contiuous}) using
(\ref{eq:numerov_z_laplacian_3}): 
\begin{equation}
\left(\alpha\beta L_{z}\right)Y_{i}(t_{k})=2\Psi_{i}(t_{k})-\left(1+\alpha V_{i}\right)Y_{i}(t_{k})+\alpha\beta\frac{\Delta z^{2}}{12}\left.\frac{\partial^{4}Y}{\partial z^{4}}\right|_{i}\label{eq:numerov_z_discrete}
\end{equation}
Now, we evaluate the error term with $\partial_{z}^{4}Y_{i}(t_{k})=L_{z}(\partial_{z}^{2}Y_{i}(t_{k}))+O(\Delta z^{2})$
also using the right hand side of (\ref{eq:numerov_z_contiuous})
to get to the result of the form 
\begin{equation}
\left(1+\alpha V_{i}+\alpha\beta L_{z}+\frac{\Delta z^{2}}{12}\cdot L_{z}\left(1+\alpha V_{i}\right)\right)Y_{i}(t_{k})=2\Psi_{i}(t_{k})+\frac{\Delta z^{2}}{6}L_{z}\Psi_{i}(t_{k}).
\end{equation}
Restoring the equations for $\Psi_{i}(t_{k+1})$ we arrive at the
Numerov-extended Crank-Nicolson algorithm as 
\begin{equation}
\left(1+\alpha V_{i}+\alpha\beta L_{z}+\frac{\Delta z^{2}}{12}L_{z}\left(1+\alpha V_{i}\right)\right)\Psi_{i}(t_{k+1})=\left(1-\alpha V_{i}-\alpha\beta L_{z}+\frac{\Delta z^{2}}{12}L_{z}\left(1-\alpha V_{i}\right)\right)\Psi_{i}(t_{k}).\label{eq:numerov_z_pade_2}
\end{equation}
It is interesting to note that by reverse engineering from the Padé-form,
we get a discrete Hamiltonian of the form 
\begin{equation}
\widetilde{H}_{z,i}=V_{i}+\beta L_{z}-i\frac{\Delta z^{2}}{6\Delta t}L_{z}+\frac{\Delta z^{2}}{12}L_{z}V_{i}.\label{eq:numerov_z_hamiltonian}
\end{equation}
The extra terms in (\ref{eq:numerov_z_hamiltonian}) improve the solution
to fifth order in $\Delta z$, however, one should note that (i) $L_{z}(V_{i}\Psi_{i}(t_{k}))$
is part of $\widetilde{H}_{z}\Psi(t_{k})$, requiring the potential
to be continuously differentiable, (ii) $\widetilde{H}_{z}$ is no
longer strictly self-adjoint.

In (\ref{eq:numerov_z_pade_2}) we have arrived at a tridiagonal system
of linear equations of the form 
\begin{equation}
\begin{bmatrix}b_{0} & c_{0} & 0 & 0 & \cdots & 0\\
a_{1} & b_{1} & c_{1} & 0 & \cdots & 0\\
0 & a_{2} & b_{2} & c_{2} & \cdots & 0\\
\vdots & \vdots & \ddots & \ddots & \ddots & \vdots\\
0 & 0 & \cdots & a_{N_{z}-1} & b_{N_{z}-1} & c_{N_{z}-1}\\
0 & 0 & \cdots & 0 & a_{N_{z}} & b_{N_{z}}
\end{bmatrix}\begin{bmatrix}\Psi_{0}(t_{k+1})\\
\Psi_{1}(t_{k+1})\\
\Psi_{2}(t_{k+1})\\
\vdots\\
\Psi_{N_{z}-1}(t_{k+1})\\
\Psi_{N_{z}}(t_{k+1})
\end{bmatrix}=\begin{bmatrix}y_{0}\\
y_{1}\\
y_{2}\\
\vdots\\
y_{N_{z}-1}\\
y_{N_{z}}
\end{bmatrix}\label{eq:numerov_z_matrix_full}
\end{equation}
which after forward Gaussian-elimination reads 
\begin{equation}
\begin{bmatrix}\tilde{b}_{0} & c_{0} & 0 & 0 & \cdots & 0\\
0 & \tilde{b}_{1} & c_{1} & 0 & \cdots & 0\\
0 & 0 & \tilde{b}_{2} & c_{2} & \cdots & 0\\
\vdots & \vdots & \ddots & \ddots & \ddots & \vdots\\
0 & 0 & \cdots & 0 & \tilde{b}_{N_{z}-1} & c_{N_{z}-1}\\
0 & 0 & \cdots & 0 & 0 & \tilde{b}_{N_{z}}
\end{bmatrix}\begin{bmatrix}\Psi_{0}(t_{k+1})\\
\Psi_{1}(t_{k+1})\\
\Psi_{2}(t_{k+1})\\
\vdots\\
\Psi_{N_{z}-1}(t_{k+1})\\
\Psi_{N_{z}}(t_{k+1})
\end{bmatrix}=\begin{bmatrix}\tilde{y}_{0}\\
\tilde{y}_{1}\\
\tilde{y}_{2}\\
\vdots\\
\tilde{y}_{N_{z}-1}\\
\tilde{y}_{N_{z}}
\end{bmatrix}.\label{eq:numerov_z_matrix_upper}
\end{equation}
In the coefficient matrix of (\ref{eq:numerov_z_matrix_full}) and
(\ref{eq:numerov_z_matrix_upper}) we denoted

\begin{equation}
a_{i}=\begin{cases}
1+\alpha V_{i-1}+12\alpha\beta/\Delta z^{2} & \text{ if }i=1,\dots,N_{z}\end{cases},\label{eq:numerov_z_matrix_a}
\end{equation}
\begin{equation}
c_{i}=\begin{cases}
1+\alpha V_{i+1}+12\alpha\beta/\Delta z^{2} & \text{ if }i=0,\dots,N_{z}-1\end{cases},\label{eq:numerov_z_matrix_c}
\end{equation}
\begin{equation}
b_{i}=\begin{cases}
10+10\alpha V_{i}-24\alpha\beta/\Delta z^{2} & \text{ if }i=0,\dots,N_{z},\end{cases}\label{eq:numerov_z_matrix_b}
\end{equation}
\begin{equation}
\tilde{b}_{i}=\begin{cases}
b_{i} & \text{ if }i=0,\\
b_{i}-(a_{i}/\tilde{b}_{i-1})c_{i-1} & \text{ if }i=1,\dots,N_{z}.
\end{cases}\label{eq:numerov_z_matrix_b_m}
\end{equation}
The right hand side and the solution of (\ref{eq:numerov_z_matrix_upper})
are given by the following expressions:

\begin{equation}
\tilde{y}_{i}=\begin{cases}
(20-b_{i})\Psi_{0}+(2-c_{i})\Psi_{1} & \text{ if }i=0,\\
(2-a_{i})\Psi_{i-1}+(20-b_{i})\Psi_{i}+(2-c_{i})\Psi_{i+1}-(a_{i}/\tilde{b}_{i-1})\tilde{y}_{i-1} & \text{ if }i=1,\dots,N_{z}-1,\\
(2-a_{i})\Psi_{N_{z}-1}+(20-b_{i})\Psi_{N_{z}}-(a_{i}/\tilde{b}_{i-1})\tilde{y}_{N_{z}-1} & \text{ if }i=N_{z},
\end{cases}\label{eq:numerov_z_matrix_rhs}
\end{equation}
\begin{equation}
\Psi_{i}(t_{k+1})=\begin{cases}
\tilde{y}_{N_{z}}/\tilde{b}_{N_{z}} & \text{ if }i=N_{z},\\
\left(\tilde{y}_{i}-c_{i}\Psi_{i+1}(t_{k+1})\right)/\tilde{b}_{i} & \text{ if }i=N_{z}-1,\dots,0.
\end{cases}\label{eq:numerov_z_result}
\end{equation}
This completes the 1D propagation method.

Now let us return to our original 2D problem. Because the discrete
Hamiltonian $H_{z}$ is independent from the value of $\rho$, the
corresponding coefficient matrix of (\ref{eq:numerov_z_matrix_upper})
is the same for each $j$-line. This means that we need to do the
forward elimination (\ref{eq:numerov_z_matrix_b_m}) only once, then
it is sufficient to perform only the forward (\ref{eq:numerov_z_matrix_rhs})
and the backward substitution (\ref{eq:numerov_z_result}) steps for
each $j$-line in order to acquire the solution. This yields a factor
of 2 speedup in the evaluation of $e^{-i\Delta t\, H_{z}}$ with three
point finite differences, if we have many lines to propagate and cache
the appropriate variables. This latter can also be viewed as an LU
decomposition based optimization \cite{BOOK_NUMERICAL_RECIPIES_2007}.

\section{Analytical solution of the forced harmonic oscillator\label{sub:forced_harmonic_oscillator}}

The TDSE for the axially symmetric (i.e. $m=0$) 3D forced harmonic
oscillator (FHO) in cylindrical coordinates is the following: 
\begin{equation}
i\frac{\partial}{\partial t}\Psi(z,\rho,t)=\left[\beta\frac{\partial^{2}}{\partial z^{2}}+\beta\frac{\partial^{2}}{\partial\rho^{2}}+\frac{\beta}{\rho}\frac{\partial}{\partial\rho}+\frac{1}{2}\mu\omega_{0}^{2}(z^{2}+\rho^{2})+f(t)z\right]\Psi(z,\rho,t)\label{eq:tdse_3d_fho}
\end{equation}
where $\beta=-1/2\mu$. Using the separability in $z$ and $\rho$
coordinates, we can reduce this problem to a one dimensional time-dependent
one, which was solved by K. Husimi et. al. \cite{husimi1953miscellanea}.
Then, the analytical time-dependent wave function is of the form 
\begin{equation}
\Psi^{{\rm A}}(z,\rho,t)=\chi(z-\xi(t),\rho,t)\exp\left(i\mu(z-\xi(t))\dot{\xi}(t)+i\int_{0}^{t}\mathcal{L}(t'){\rm d}t'\right),\label{eq:solution_3d_fho}
\end{equation}
where $\chi(z,\rho,t)$ is a solution of the 3D field-free quantum
harmonic oscillator problem with axial symmetry. We define it to be
the ground state of the field-free problem (H-000): 
\begin{equation}
\chi_{000}(z,\rho,t)=\left[\frac{\mu\omega_{0}}{\pi}\right]{}^{3/4}e^{-\frac{1}{2}\mu\omega_{0}\left(z^{2}+\rho^{2}\right)}e^{-i\frac{3}{2}\omega_{0}t}.
\end{equation}
In formula (\ref{eq:solution_3d_fho}) the symbol $\mathcal{L}(t)$
denotes the Lagrangian of the corresponding classical system: 
\begin{equation}
\mathcal{L}(t)=\frac{1}{2}\mu\dot{\xi}(t)^{2}-\frac{1}{2}\mu\omega_{0}^{2}\xi(t)^{2}-f(t)\xi(t),
\end{equation}
and $\xi(t)$ is the solution of the initial value problem 
\begin{equation}
\ddot{\xi}(t)=f(t)/\mu-\omega_{0}^{2}\xi(t),\text{ with }\{\xi(0)=0,\dot{\xi}(0)=0\}.
\end{equation}
We set $f(t)=F\sin\omega_{{\rm F}}t$, then the previous equation
is that of the forced harmonic oscillator has the solution:

\begin{equation}
\xi(t)=\frac{F/\mu}{\omega_{{\rm F}}^{2}-\omega_{0}^{2}}\left[\sin\omega_{{\rm F}}t-\frac{\omega_{{\rm F}}}{\omega_{0}}\sin\omega_{0}t\right].
\end{equation}

These formulae define the analytical solution that we use in Section
\ref{sub:test_3d_harmonic} as one of our test cases. There, in Figure
\ref{fig:test_3d_harmonic_support}, we also plot $\xi(t)$, which
is actually the expectation value of the coordinate $z$, and the
different terms of the phase in (\ref{eq:solution_3d_fho}). 
\end{document}